\documentclass[aps,11pt,prd,groupedaddress,nofootinbib,notitlepage,eqsecnum,preprintnumbers]{revtex4-2}

\usepackage[utf8]{inputenc}
\usepackage{graphicx}
\usepackage{amsmath, amssymb}
\usepackage{mathtools}
\usepackage{physics}
\usepackage[english]{babel}
\usepackage[colorlinks=true, allcolors=blue]{hyperref}
\usepackage{url}
\usepackage{cleveref}
\usepackage{xcolor}
\usepackage[normalem]{ulem}
\usepackage{soul}
\usepackage{bbold}
\usepackage[lofdepth,lotdepth,caption=false]{subfig}
\usepackage{tabularx}
\usepackage{multirow}

\usepackage{pifont}
\usepackage{fontawesome5}

\newcommand{\pathology}{\textcolor{red!75!black}{\faTimesCircle}}      
\newcommand{\warning}{\textcolor{orange!90!black}{\faExclamationTriangle}} 
\newcommand{\healthy}{\textcolor{green!60!black}{\faCheckSquare}}      

\graphicspath{{./figures/}}

\def\calG{\mathcal{G}}
\def\calGH{\mathcal{G}_{\rm H}}
\def\calF{\mathcal{F}}
\def\calL{\mathcal{L}}
\def\calM{\mathcal{M}}
\def\Mpl{M_{\rm Pl}}

\begin{document}
\preprint{YITP-26-137}

\title{\textit{All Quiet on the Vector Front:} \\ Scalar Growth and Perturbative Control in Horndeski--Weyl-squared Gravity}

\author{\textsc{Antonio De Felice$^{a}$}}
    \email{{antonio.defelice}@{yukawa.kyoto-u.ac.jp}}
\author{\textsc{Guillem Dom\`enech$^{b,c}$}}
    \email{{guillem.domenech}@{itp.uni-hannover.de}}
\author{\textsc{Jan Tränkle$^{b,d,e}$}}
\email{{jan.traenkle}@{itp.uni-hannover.de}}

\affiliation{$^a$
Center for Gravitational Physics and Quantum Information,
Yukawa Institute for Theoretical Physics, Kyoto University, 606-8502, Kyoto, Japan}
\affiliation{$^b$Institute for Theoretical Physics, Leibniz University Hannover, Appelstraße 2, 30167 Hannover, Germany}
\affiliation{$^c$Max Planck Institute for Gravitational Physics, Albert Einstein Institute, 30167 Hannover, Germany}
\affiliation{$^d$Dipartimento di Fisica e Astronomia ``G. Galilei'', Università degli Studi di Padova, via Marzolo 8, I-35131 Padova, Italy}
\affiliation{$^e$INFN, Sezione di Padova, via Marzolo 8, I-35131 Padova, Italy}

\begin{abstract}
    Higher-order curvature corrections to general relativity arise naturally in many fundamental theories and provide phenomenologically interesting low-energy effective field theories. However, higher-derivative terms can introduce ghost modes, which call the validity of such models into question. Here, we thoroughly investigate the classical linear stability of the most general second-order scalar-tensor theory, namely the Horndeski theory of gravity, augmented by a Weyl-squared term in the action. We consider shift-symmetric de Sitter and power-law FLRW backgrounds and study the evolution of linear perturbations. Vector perturbations are free of gradient and tachyonic instabilities for a positive Weyl-induced mass squared. Tensor perturbations necessarily contain a ghost; their superhorizon branches are bounded, while a frozen-background calculation identifies a separate finite-momentum stability condition. The scalar sector instead contains a growing long-wavelength branch for all expanding backgrounds we consider, which is inherited by at least one of the gauge-invariant Bardeen potentials. We show that this growth enters the Bardeen potentials through an inverse-gradient factor $a^2/k^2$: the dynamical scalar fields remain bounded on the inflationary backgrounds, the direct contribution of the growing shear to the clock-matched tidal field cancels, and the electric part of the Weyl tensor decays. We discuss the implications of our results for perturbative control and for the phenomenological viability of theories including quadratic curvature corrections.
\end{abstract}

\maketitle

\section{Introduction} \label{sec:Introduction}
The paradigm of cosmic inflation provides an elegant mechanism to explain the large-scale homogeneity and isotropy of the Universe, as well as the nearly scale-invariant spectrum of primordial perturbations that seed the observed large-scale structure \cite{Starobinsky:1980te, Kazanas:1980tx, Guth:1980zm, Linde:1981mu, Albrecht:1982wi,Liddle:2000cg, Martin:2013tda}. Among the many proposed models, the Starobinsky model \cite{Starobinsky:1980te}, driven by a quadratic Ricci scalar $R^2$, remains one of the best-fit models to current CMB observations \cite{Planck:2018vyg,Planck:2018jri}.

From the perspective of effective field theory (EFT) and the ultraviolet (UV) completion of gravity, the Starobinsky model arises naturally as the leading-order correction to the Einstein–Hilbert action in a curvature expansion of the gravitational effective action $\Gamma[g_{\mu\nu}]$ \cite{Donoghue:1994dn, Burgess:2003jk}. At low energies, gravity is described by a derivative expansion in powers of curvature invariants; the zeroth-order term is the Einstein–Hilbert action with cosmological constant, while the first corrections are of order $\mathcal{O}(R^2)$. However, at the same order in this expansion, one generically expects the presence of all quadratic curvature invariants compatible with diffeomorphism invariance. Given the Gauss–Bonnet identity, the independent quadratic invariants are $R^2$ and the squared Weyl tensor $C^2 \coloneqq C^{\alpha\beta\mu\nu} C_{\alpha\beta\mu\nu}$. The Weyl term is therefore a generic prediction of UV completions of gravity, including asymptotically safe gravity, loop quantum gravity, and string theory \cite{Bambi:2023jiz, Pawlowski:2023dda}, and the same conclusion is reached independently from perturbative attempts to renormalize GR itself, where loop corrections are well known to induce a tower of higher-order curvature counterterms \cite{tHooft:1974toh,Goroff:1985sz,Goroff:1985th}.
In such UV completions, the dimensionless couplings of the quadratic invariants are generically expected to be of order unity; in asymptotically safe gravity, for instance, the quadratic-curvature couplings approach interacting fixed points with $\mathcal{O}(1)$ values \cite{Codello:2006in, Benedetti:2009rx}.
For a Weyl coupling $\alpha$, the mass of the additional spin-2 mode is parametrically $m_2\sim \Mpl/\sqrt{|\alpha|}$ in the weak-curvature limit \cite{Stelle:1977ry,Salvio:2018crh}.
For an $\mathcal{O}(1)$ Weyl coupling $\alpha$, the mass $m_2$ is of the order of the Planck scale. Whether the extra modes must be retained during inflation depends on their mass relative to the EFT cutoff and the scales being probed.

The Weyl-squared term is known to introduce a ghost degree of freedom (d.o.f.) arising from higher-derivative field equations. This leads to the well-known Ostrogradsky instability, which generically appears in higher-derivative theories \cite{Ganz:2020skf}.
While gradient and tachyonic instabilities threaten the viability of a classical theory, ghost instabilities are usually considered fatal for the quantized theory, as ghost particles are associated with negative energy and can therefore be created from the vacuum in a runaway cascade; see \cite{Sbisa:2014pzo} for a review. However, this paradigmatic assumption has recently been questioned, and there have been several proposals for how to deal with the Ostrogradsky ghost in a quantum theory, see e.g.~\cite{Bender:2007wu, Modesto:2015ozb, Anselmi:2017ygm, Anselmi:2018kgz, Donoghue:2019fcb, Platania:2020knd, Donoghue:2021eto, Deffayet:2021nnt, Glavan:2024cfs,Deffayet:2025lnj,Held:2025fii,Deffayet:2026cnu,Deffayet:2026uoe}.
From a quantum gravity viewpoint, the emergence of ghosts at quadratic order in curvature may be interpreted as an artifact due to the truncation of the perturbative expansion of the underlying, non-perturbative theory \cite{Platania:2020knd,Platania:2022gtt}. In this case, one would expect that classically stable solutions should exist despite the presence of ghost modes.

The quadratic curvature theory
\begin{align}
    \calL_{\rm QG}= \sqrt{-g}\left(\frac{\Mpl^2}{2}R + \beta R^2 - \alpha C^2\right) \,,
\end{align}
first studied by Stelle \cite{Stelle:1976gc,Stelle:1977ry}, has been shown to be perturbatively renormalizable, but the appearance of the Weyl ghost raises concerns about classical and quantum stability \cite{Salvio:2018crh}.
Indeed, on a flat Minkowski background, the spectrum of quadratic gravity consists of a massless graviton together with a massive scalar and a massive spin-2 ghost, but (in the absence of additional matter fields) no Laplacian instabilities arise \cite{Hindawi:1995an, Bogdanos:2009tn, Hinterbichler:2015soa}; the ghost modes are thus ``soft" in the sense of \cite{Smilga:2004cy}.
Furthermore, it was found that classical stability in the presence of the Weyl term can also be achieved during the matter-dominated era, for a suitable range of the coupling \cite{Sutton:2025ixz}.
Whether this benign behavior persists on an inflating background is therefore an important, non-trivial question which has received significant attention \cite{Clunan:2009er, Deruelle:2010kf,Ivanov:2016hcm, Salvio:2017xul,Anselmi:2020lpp,DeFelice:2023psw,Palomares:2026iwt}.
Ref.~\cite{Deruelle:2010kf} investigated the stability of an inflating background with a minimally coupled scalar field and the Weyl-squared term, finding that the scalar metric perturbations diverge in the Newtonian gauge but remain finite in the comoving slicing, and moreover that the vector perturbations can no longer be neglected once the Weyl term is included.

Whether this apparent gauge-dependence reflects a physical instability or merely represents a gauge artifact has become the subject of a recent debate: in Ref.~\cite{DeFelice:2023psw} it was shown that, while the vector and tensor ghosts remain soft during inflation (meaning that perturbations decay or approach constants after Hubble radius crossing), the scalar sector harbors a purely classical superhorizon instability. The two gauge-invariant Bardeen potentials grow exponentially after Hubble exit, which was interpreted there as rendering the homogeneous Friedmann-Lema\^itre-Robertson-Walker (FLRW) background unviable. The Weyl ghost in the scalar sector was therefore argued to be of the hard type in the Starobinsky background.
Ref.~\cite{Palomares:2026iwt} subsequently studied scalar perturbations in quadratic gravity in the Einstein frame, casting the dynamics in terms of several gauge-invariant variables. In the superhorizon de Sitter limit they find exponentially growing branches in some variables, while others approach a constant. After examining the metric and scalar-field perturbations in Newtonian, flat, and comoving gauges, they argue that the Newtonian-gauge growth is a gauge artifact, in line with the interpretations put forward in \cite{Ivanov:2016hcm,Salvio:2017xul}. These discussions raise a natural question: does the scalar instability -- and the ambiguity surrounding its gauge (in)dependence -- persist once the inflationary background and gravitational theory are generalized beyond the Starobinsky model?

A well-motivated and maximally general framework for this generalization is Horndeski gravity \cite{Horndeski:1974wa,Kobayashi:2019hrl}, the most general scalar-tensor theory in four spacetime dimensions whose field equations are second order in derivatives. Horndeski theory encompasses a wide class of inflationary models, including canonical slow-roll inflation, $k$-inflation, and the Starobinsky model itself. Its four free functions $G_{2,3,4,5}(\phi, X)$ parameterize a large landscape of inflationary backgrounds, and one may hope that for appropriate choices of these functions the Weyl ghost is rendered soft.
Such classical stability is a natural first diagnostic before applying the so-called fakeon prescription \cite{Anselmi:2017ygm,Anselmi:2018kgz,Anselmi:2020lpp} to the Weyl ghost: promoting the ghost pole to a purely virtual particle removes it from the physical spectrum in a way consistent with unitarity, potentially rendering the theory quantum-mechanically viable. The fakeon prescription, however, also restricts the admissible physical data and modifies the classical response \cite{Anselmi:2018bra,Anselmi:2025uda}. The analysis of the unprojected local equations presented here is therefore a diagnostic of the local action rather than a test of a specific fakeon implementation. Understanding the classical behavior of the local theory is nevertheless a natural first step toward a consistent EFT describing inflation at quadratic order in curvature.

In this paper, we systematically investigate linear scalar, vector, and tensor perturbations in Horndeski plus Weyl-squared theory on expanding FLRW backgrounds. We find that the vector and tensor sectors exhibit the same qualitative behavior as in Stelle's theory: despite the presence of ghost modes, the classical perturbations remain bounded, provided the Horndeski functions satisfy the conditions derived below. The scalar sector is more subtle. For both classes of expanding backgrounds we consider -- shift-symmetric de Sitter backgrounds and general power-law solutions -- we find a superhorizon growing branch analogous to, and generalizing, the one found in \cite{Deruelle:2010kf,Ivanov:2016hcm,Salvio:2017xul,DeFelice:2023psw,Palomares:2026iwt}.
In particular, we find that the growing branch persists throughout the classes of backgrounds considered here and, although its manifestation depends on the gauge -- in unitary gauge the shear potential grows, while in Newtonian gauge at least one of the gauge-invariant Bardeen potentials grows -- the growth of the Bardeen potentials cannot be eliminated by a coordinate choice.
At the same time, we show that this growth enters the Bardeen potentials through an inverse-gradient factor $a^2/k^2$: the dynamical scalar fields remain bounded on the inflationary backgrounds, the direct contribution of the growing shear to the clock-matched tidal field cancels, and the electric part of the Weyl tensor decays.
Our results thus show that, within the classes we consider, no inflationary background of the Horndeski plus Weyl-squared theory has bounded Bardeen potentials at linear order, and that the growing branch limits the domain of validity of the Newtonian-gauge weak-metric expansion. Whether it constitutes a physical instability is a separate question, which we discuss in \cref{sec:discussion}.
\medskip

The rest of this paper is organized as follows. In \cref{sec:Horndeski-Weyl-theory} we introduce the action of our Horndeski--Weyl-squared model and the two classes of inflationary backgrounds -- the shift-symmetric de Sitter case and the power-law solutions -- that we consider throughout. \Cref{sec:background-eqs} derives the corresponding background equations of motion. In \cref{sec:stability-pert} we set up the metric perturbations, define the ghost, gradient, and tachyonic instabilities we check for, and suggest several criteria for classical and phenomenological viability which a given theory should satisfy. We then study vector perturbations in \cref{sec:vector-perturbations}, finding that the Weyl term renders them dynamical but classically stable (under some conditions), and tensor perturbations in \cref{sec:tensor-perturbations}, where we show that a ghost mode is unavoidable yet the classical perturbations remain bounded under conditions on the Horndeski functions. \Cref{sec:scalar-perturbations} contains our main result: after establishing the no-ghost and no-gradient-instability conditions, we find that the scalar sector harbors a superhorizon growing branch in the shear potential (in unitary gauge), which is inherited by (at least one of) the gauge-invariant Bardeen potentials for both classes of background. We discuss the implications and interpretation of our results in \cref{sec:discussion}.
Finally, \cref{sec:conclusion} summarizes our results and discusses the physical significance of these growing modes and their implications for the viability of the theory. The appendices contain supporting material: the second-order action of GR, the tensor kinetic diagonalization, the geodesic-deviation and tidal identities, the closed Bardeen equations, and the finite-momentum two-scalar reduction.

\section{Horndeski theory with a Weyl-squared term}\label{sec:Horndeski-Weyl-theory}
We consider Horndeski theory augmented with the square of the Weyl tensor,
which is given by the action
\begin{equation} \label{eq:Action-Horndeski-Weyl}
    S = \int \text{d}^4x \sqrt{-g} \left( \frac{\Mpl^2}{2} R +  \calL_H - \alpha C^2 \right) \,.
\end{equation}
Before explaining our notation, let us clarify that we interpret \cref{eq:Action-Horndeski-Weyl} as a truncation of a gravitational EFT. We denote by $\Lambda_{\rm EFT}$ the scale suppressing the higher-curvature operators that have been omitted, schematically $R^3/\Lambda_{\rm EFT}^2$, $R\,C^2/\Lambda_{\rm EFT}^2$, and so on \cite{Donoghue:1994dn,Burgess:2003jk}. The scale $\Lambda_{\rm EFT}$ is conceptually distinct from the mass scale $\calM$ associated with the additional mode generated by the Weyl-squared term (given later in \cref{eq:M2-definition}). Unless a UV completion or a matching calculation specifies the Wilson coefficients, the two scales should not be identified. Whenever an EFT interpretation is invoked, the background curvature and the physical frequencies/momenta must remain parametrically below $\Lambda_{\rm EFT}$.
If the mass scale lies at or above the EFT cutoff, the corresponding pole should not be retained as an independent low-energy degree of freedom; it is instead integrated out (or removed perturbatively by order reduction) \cite{Appelquist:1974tg}. The local higher-derivative equations can be studied as a classical theory in their own right. Interpreting their additional solutions as controlled low-energy EFT modes requires the stronger assumption that the pole and its interactions lie within a valid effective description. We keep this distinction explicit throughout.

Now, in \cref{eq:Action-Horndeski-Weyl}, $C^2 \coloneqq C^{\alpha\beta\mu\nu}C_{\alpha\beta\mu\nu}$ is the Weyl tensor squared\footnote{The Weyl-squared term can be expressed in terms of the Ricci scalar and tensor as
\begin{equation} \label{eq:Weyl2-R-GB}
    C^2 = 2 R^{\mu\nu} R_{\mu\nu} - \frac{2}{3} R^2 + \calG_{\rm GB} \,,
\end{equation}
with the Gauss-Bonnet term $\calG_{\rm GB} = R^{\alpha\beta\mu\nu} R_{\alpha\beta\mu\nu} - 4 R^{\mu\nu} R_{\mu\nu} + R^2$. In fact, $\calG_{\rm GB}$ constitutes a topological invariant (the Euler characteristic) when integrated over the spacetime manifold, and thus does not contribute to the equations of motion, so that we can always replace $ C^2 = 2 R^{\mu\nu} R_{\mu\nu} - \frac{2}{3} R^2$ in the action.}, $\alpha$ is its dimensionless coupling, $\Mpl\coloneqq 1/\sqrt{8\pi G_N}$ is the reduced Planck mass, and the Lagrangian density of Horndeski gravity is \cite{Horndeski:1974wa,Kobayashi:2019hrl}
\begin{align}
    \calL_H = & \,G_2(\phi,X) - G_3(\phi,X)\Box \phi + G_4(\phi,X) R + G_{4,X} \left((\Box \phi)^2 - \phi^{\mu\nu}\phi_{\mu\nu}\right) \nonumber \\
    & + G_5(\phi,X) G^{\mu\nu}\phi_{\mu\nu} - \frac{1}{6} G_{5,X} \left((\Box \phi)^3 - 3 \Box \phi \phi^{\mu\nu}\phi_{\mu\nu} + 2 \phi_{\mu\nu}\phi^{\nu\lambda}\phi_\lambda^\mu \right) \,,
    \label{eq:Horndeski-Lagrangian}
\end{align}
where the kinetic term of the Horndeski scalar degree of freedom $\phi$ is
\begin{align}
    X \coloneqq - \frac{1}{2}\nabla_\mu \phi \nabla^\mu\phi \,.
\end{align}
For the sake of clarity of notation, we will suppress the arguments and only write $G_i(\phi,X)\equiv G_i$, with $i\in[2,5]$, in the following.
In \cref{eq:Horndeski-Lagrangian} we also introduced the notation
\begin{align}
    \phi_\mu \coloneqq \nabla_\mu \phi \,,\quad
    \phi_{\mu\nu} \coloneqq \nabla_\mu \nabla_\nu \phi \,,\quad \text{and}
    \quad \Box \phi \coloneqq \nabla^\mu \nabla_\mu \phi \,,
\end{align}
as well as $f_x \coloneqq \partial f/ \partial x$ for a generic function $f\equiv f(x)$.

Horndeski theory is, by construction, the most general scalar-tensor theory in four dimensions with second order field equations \cite{Horndeski:1974wa}. In particular, it also encompasses all models where a single canonical scalar field $\phi$ with potential $V(\phi)$ drives inflation, namely when $G_2(\phi,X) = X-V(\phi)$ and $G_{3,4,5}=0$.
As such it also includes the Starobinsky $R^2$ model, as well as more general $f(R)$ theories, which can be written as general relativity (GR) with a minimally coupled scalar field \cite{Whitt:1984pd}.
In \cref{eq:Action-Horndeski-Weyl} we chose to keep the Einstein-Hilbert term separate, but it could also be absorbed in the $G_4$-term. In other words, $G_4$ leads to a field-dependent modification of the Newton constant (or Planck mass), $\tilde{M}_{\rm Pl}^2 \coloneqq \Mpl^2 + 2 G_4$.
In the context of $k$-inflation, the first term is often denoted as $G_2 = \mathcal{P}$ \cite{Armendariz-Picon:1999hyi}, and in the so-called G-inflation model it is written as $G_2 = K$ \cite{Kobayashi:2010cm}.
Inflation in the full Horndeski theory has been studied in \cite{Kobayashi:2011nu}, where it was termed ``Generalized G-inflation".

Let us note that, when considering Horndeski theory to explain the accelerated expansion at late times (usually attributed to dark energy), the terms $G_{4,X}$, $G_5$ and $G_{5,X}$, which lead to a modified propagation speed of tensor perturbations $c_t\neq 1$, are essentially ruled out by the gravitational wave event GW170817, which tightly constrained the speed of gravitational waves to be close to the speed of light, $c_t \simeq 1$ \cite{LIGOScientific:2017zic,Kase:2018aps}. However, in the early Universe -- and in particular during inflation -- these terms may still play a significant role. For example, there are models for which $c_T\to1$ is a late time attractor \cite{Amendola:2018ltt}.
Ref.~\cite{Kase:2018aps} also provides an overview of classes of models that are included in the Horndeski Lagrangian \cref{eq:Horndeski-Lagrangian}.

In order to enable practical computations, we will have to make some assumptions about the functional form of the free Horndeski functions $G_i(\phi,X)$. We present the two general subclasses we consider separately below.

\subsection{Shift-symmetric case in de Sitter background  \label{sec:shiftsymmetrydS}}
The first subclass of Horndeski theory that we consider is defined by the condition
\begin{align}
    G_{i,\phi}\equiv0 \,, \quad \text{and} \quad X=\frac12 \dot\phi^2\equiv\text{const.} \,,
\end{align}
such that $G_i=G_i(X)$ and their background values are constant. Their derivatives with respect to $X$ need not vanish. Such a setup is inspired by $k$-inflation, where only the (non-canonical) kinetic part is responsible for driving inflation \cite{Armendariz-Picon:1999hyi}.
In this setup, we will also take the quasi-de-Sitter limit for the inflationary dynamics, i.e., we will assume
\begin{align}
    |\dot H|\ll H^2 \,, \quad \text{and} \quad H\simeq H_I\equiv\text{const.}\,,
\end{align}
such that the scale factor $a$ of the FLRW spacetime grows exponentially with time,
\begin{align}
    a(t) = a_* e^{H_I(t-t_*)} \,.
\end{align}
Here we introduced the Hubble parameter $H\coloneqq\dot{a}/a$, and a dot ( $\dot{}$ ) denotes a derivative with respect to cosmic time $t$.
Under these assumptions, we can explicitly solve the perturbation equations. In practice, the explicit solutions below use the exact limit $H=H_I$ with constant $X>0$, so that $\dot\phi\neq0$ defines a regular uniform-field slicing.

\subsection{General power-law solutions \label{sec:powerlawsolutions}}
As the second subclass of theories, we consider those that lead to a power-law expansion of the background spacetime with a scale factor that evolves as
\begin{align}
    a=a_{*}\left(\frac{t}{t_{*}}\right)^{p} \,, \quad \text{with} \quad H=\frac{p}{t} \,.
\end{align}
Such a power-law expansion can be interpreted as if the Universe were filled with a fluid with a constant effective equation of state parameter $w$, which is related to the power-law index by
\begin{align}\label{eq:relation-p-w}
    p=\frac{2}{3(1+w)} \,.
\end{align}
Let us emphasize, however, that these solutions allow for any value of $p$ a priori.
For convenience, we list some example values in \cref{tab:p_vs_w}. Accelerated expansion (as is required for inflation) is achieved for $p>1$; the limiting case $p=1$ corresponds to coasting expansion with $\ddot a=0$. Although $p<0$ is also possible, see e.g.~\cite{Domenech:2015qoa}, for simplicity we focus on cases with $p>0$.
%
\begin{table*}[t]
{\def\arraystretch{1.75}
\centering
    \begin{tabular}{|c||c|c|c|c|c|c|}
     \hline
     $w$ & $-1$ & $-2/3$ & $-1/3$ & $0$ & $1/3$ & $1$ \\
     \hline
     $p$ & $\infty$ & $2$ & $1$ & $2/3$ & $1/2$ & $1/3$ \\
     \hline
    \end{tabular}}
    \caption{Some values of the power-law exponent $p$ and corresponding values of the (effective) equation of state parameter $w$ from the relation \cref{eq:relation-p-w}.}
    \label{tab:p_vs_w}
\end{table*}
%

While satisfying the power-law behavior of $a$, one can also work with a field redefinition such that the scalar field $\phi$ is linear in $t$
\begin{align}
    \phi=\phi_{*}\,\frac{t}{t_{*}} \,.
\end{align}
This can be achieved by fixing the functions $G_i$ to \cite{Domenech:2024drm}
\begin{align}
G_2 & =\Lambda_{2}^{4}\left(\frac{\phi}{\phi_{*}}\right)^{-2}a_{2}(X)\,, \qquad
&&G_{3} =\Lambda_{3}\left(\frac{\phi}{\phi_{*}}\right)^{-1}a_{3}(X)\,,\\
G_{4} & =\Lambda_{4}^{2}\,a_{4}(X)\,,\qquad
&&G_{5} =\frac{1}{\Lambda_{5}}\left(\frac{\phi}{\phi_{*}}\right)a_{5}(X)\,,
\end{align}
where $X=\phi_{*}^{2}/(2t_{*}^{2})=\text{constant}$. Each $\Lambda_i$, with $i\in[2,5]$, has mass dimension 1, ensuring the correct dimensionality of the equations. For convenience, we absorb the $\Lambda_i$'s into the free functions $a_i(X)$ from now on, but we show them explicitly here so they can be easily restored later if needed. We will further normalize the scalar field $\phi$ by the reference value $\phi_*$, making it dimensionless.
\medskip

By allowing for both the quasi-de-Sitter limit in the shift-symmetric case, as well as the non-exponential, power-law accelerated expansion we expect to cover a large set of possible inflationary scenarios within the two cases we consider.

\section{Background equations}\label{sec:background-eqs}
Starting with the background equations, we consider a flat FLRW metric with the line element
\begin{align} \label{eq:FLRW-background}
    ds^2 = -N(t)^2\text{d}t^2 + a(t)^2 \text{d}\vec x^2 \,,
\end{align}
where the lapse is kept arbitrary during variation and is subsequently set to $N(t)\equiv1$, without loss of generality.
At the background level, the scalar field is homogeneous and only depends on time, $\phi\equiv\phi(t)$, so that $X= \frac{1}{2}\dot{\phi}^2$ and $\dot{X} = \dot{\phi} \ddot{\phi}$.
A variation of the action \cref{eq:Action-Horndeski-Weyl} with respect to $g^{\mu\nu}$ yields the equations of motion.
Adopting the notation of \cite{Kobayashi:2011nu}, up to an overall sign of the energy equation, we can write the $00$-component (corresponding to variation with respect to $N(t)$) of the resulting background equations as
\begin{align}
    \sum_{i=1}^5\mathcal{E}_i = 0 \,,
    \label{eq:Bkg_EoM_00}
\end{align}
where the $\mathcal{E}_i$ represent the contributions from the individual terms in \cref{eq:Action-Horndeski-Weyl}. They are given by
\begin{align}
    \mathcal{E}_1 &= 3 H^2 \Mpl^2 \,, \\
    \mathcal{E}_2 &= G_2- 2X G_{2,X} \,, \\
    \mathcal{E}_3 &= 2X\left(G_{3,\phi} -3 H \dot{\phi} G_{3,X}\right) \,, \\
    \mathcal{E}_4 &= 6 H \left(HG_4 + \dot{\phi} G_{4,\phi}-4 H X \left( G_{4,X} + X G_{4,XX}\right)+2 X \dot{\phi} G_{4,\phi X} \right) \,, \\
    \mathcal{E}_5 &= 2H^2X\left(9 G_{5,\phi} - H \dot{\phi}\left(5 G_{5,X} + 2 X G_{5,XX}\right) +6 X G_{5,\phi X}\right) \,.
\end{align}
The trace of the $ij$-component (stemming from variation with respect to $a(t)$) results in
\begin{align}
    \sum_{i=1}^5\mathcal{P}_i = 0 \,,
    \label{eq:Bkg_EoM_ij}
\end{align}
with the $\mathcal{P}_i$ given by
\begin{align}
    \mathcal{P}_1 &= \left(3 H^2 + 2 \dot{H}\right) \Mpl^2 \,, \\
    \mathcal{P}_2 &= G_{2} \,, \\
    \mathcal{P}_3 &= -2 X \left( G_{3,\phi} + \ddot{\phi} G_{3,X} \right) \,, \\
    \mathcal{P}_4 &= 2\left(3H^2+ 2\dot{H}\right) G_{4}+2 (2 H \dot{\phi}+\ddot{\phi}) G_{4,\phi} -4 \left(\left(3 H^2+2 \dot{H}\right)X+ H \dot{X}\right) G_{4,X} \nonumber \\
    &\phantom{=}+4X G_{4,\phi \phi} - 8 H X \dot{X} G_{4,XX} + 4 X(\ddot{\phi}-2 H \dot{\phi}) G_{4,\phi X} \,, \\
    \mathcal{P}_5 &= 2 \left(\left(3 H^2+2 \dot{H}\right) X + 2 H \dot{X}\right) G_{5,\phi} - 2 X \left(3 H^2 \ddot{\phi}+2 \dot{\phi} \left(H^3+H \dot{H}\right)\right)G_{5,X} \nonumber \\
    & \phantom{=}+ 4 H \dot{\phi} X G_{5,\phi \phi} - 4 H^2 \ddot{\phi} X^2 G_{5,XX} +4 H X \left(\dot{X}-H X\right) G_{5,\phi X} \,.
\end{align}

Finally, the variation of the action with respect to the scalar field leads to
\begin{align}
    \mathcal{I} - 3H \mathcal{J} - \dot{\mathcal{J}} = 0 \,,
    \label{eq:Bkg_EoM_Phi}
\end{align}
where
\begin{align}
    \mathcal{I} = & G_{2,\phi}-2X\left(G_{3,\phi\phi}+\ddot\phi G_{3,\phi X}\right)+6\left(2H^2+\dot H\right)G_{4,\phi} \nonumber \\
    & +6H\left(\dot X + 2HX\right)G_{4,\phi X} - 2H^2 X\left(3 G_{5,\phi\phi}-H\dot\phi G_{5,\phi X} \right)\,,
\end{align}
and
\begin{align}
    \mathcal{J} = \, & \dot\phi G_{2,X} + 6HXG_{3,X} - 2\dot\phi G_{3,\phi} + 6H^2\dot\phi\left(G_{4,X} + 2XG_{4,XX}\right) - 12HXG_{4,\phi X} \nonumber \\
    & +2H^3 X\left(3 G_{5,X}+2XG_{5,XX}\right) - 6 H^2\dot\phi\left(G_{5,\phi}+XG_{5,\phi X}\right) \,.
\end{align}
Only two of the equations \labelcref{eq:Bkg_EoM_00,eq:Bkg_EoM_ij,eq:Bkg_EoM_Phi} are mutually independent \cite{Kobayashi:2011nu}.

Note that the Weyl-squared term does not contribute at the background level due to its conformal invariance. Specifically, this means that $C^\alpha_{\phantom{\alpha}\beta\mu\nu}[g] \equiv C^\alpha_{\phantom{\alpha}\beta\mu\nu}[\tilde g]$ for metrics related by a conformal transformation, $\tilde g_{\mu\nu} = \Omega^2 g_{\mu\nu}$. Introducing conformal time by $\text{d}t=a\text{d}\tau$ shows that the FLRW metric \cref{eq:FLRW-background} is conformally related to the flat Minkowski metric by $g_{\mu\nu}=a^2 \eta_{\mu\nu}$, such that $C^2[g]=a^{-4}C^2[\eta]=0$. We proceed to study the cases presented in \cref{sec:shiftsymmetrydS,sec:powerlawsolutions} separately below.

\subsection{Shift-symmetric dS case}
In the shift-symmetric dS case, the background equations reduce to
\begin{align}
    0 =& \,3 H_I^2 \Mpl^2 + G_2 - 2 X G_{2,X}
        + 6 H_I^2 \left(G_4 - 4 X G_{4,X} - 4 X^2 G_{4,XX}\right) \nonumber \\
      &- 2 H_I X \dot\phi \left[3 G_{3,X} + H_I^2 \left(5 G_{5,X} + 2 X G_{5,XX}\right)\right] \,, \\
    0 =& \, G_{2,X} + 6 H_I^2 \left(G_{4,X} + 2 X G_{4,XX}\right)
        + H_I \dot\phi \left[3 G_{3,X} + H_I^2 \left(3 G_{5,X} + 2 X G_{5,XX}\right)\right] \,.
\end{align}

\subsection{Power-law solutions}
For the power-law solutions, the equations of motion can be expressed as
\begin{align}
    0 =& \, a_2 - 2 X \Big[ a_{2,X} + a_3 + 6 p X a_{3,X}
        + 6 p^2 \left(4 X^2 a_{4,XX} + 4 X a_{4,X} - a_4\right) \nonumber \\
      &\phantom{a_2 - 2 X \Big[}
        + 2 p^2 X \left(4 p X^2 a_{5,XX} + 2(5p-3) X a_{5,X} - 9 a_5\right)
        - 3 p^2 \Mpl^2 \Big] \,, \\
    0 =& \,a_2 + 2 X \left[ a_3 + p (3p-2)\left(\Mpl^2 + 2 a_4 - 4 X a_{4,X} + 2 X a_5\right)
        - 4 p^2 (2p-1) X^2 a_{5,X} \right] \,.
\end{align}

\section{Stability of perturbations}\label{sec:stability-pert}
We now turn to study perturbations of the background FLRW metric \cref{eq:FLRW-background}, which we parametrize as
\begin{equation} \label{eq:FLRW-unitary}
    ds^2 = - (1+2 \psi)\text{d}t^2 + 2 \left(a B_i+\partial_i \chi\right) \text{d}t \text{d}x^i + a^2 \left( \left( 1 + 2 \zeta \right) \delta_{ij} + 2a^{-2} \partial_i\partial_j E+  h_{ij}\right) \text{d}x^i \text{d}x^j \,,
\end{equation}
where $\psi$ denotes the perturbation of the lapse, $\zeta$ is the curvature perturbation, $E$ is the scalar spatial metric perturbation, $B_i$ denotes a vector perturbation (which does not propagate in GR, but becomes dynamical upon including the Weyl-term), and $h_{ij}$ are the tensor perturbations, which constitute gravitational waves on deep-subhorizon scales. The vectors are divergence-free (transverse), meaning $\partial^i B_i = 0$, and the tensor perturbations satisfy the transverse-traceless conditions, i.e.~$\partial^i h_{ij}=0$ and $\delta^{ij} h_{ij}=0$. In the calculations below, we fix the scalar spatial gauge by setting $E=0$. Keeping $E$ explicit in \cref{eq:FLRW-unitary} is nevertheless useful when comparing different gauge choices.
For general reviews and introductions to cosmological perturbation, we refer the reader to \cite{Kodama:1984ziu,Mukhanov:1990me,Malik:2008im,Baumann:2022mni}.

Note that here we normalize the perturbation variables so that spatial displacements and derivatives in terms of the comoving coordinates $\vec x$ are accompanied by the corresponding powers of the scale factor, i.e.~they appear as $a\text{d}x$ and $a^{-1}\partial_i$. Differently normalized fields, such as e.g.~$\widetilde{\chi}\coloneqq \chi/a$, can therefore have a different late-time scaling simply because of the background expansion. Such a rescaling alone does not determine whether a growing solution is physically pathological. In the conventions of Ref.~\cite{Palomares:2026iwt}, which we denote with a subscript ``PZK" from now on, the scalar metric perturbations are related to ours by
\begin{equation}
 A_{\rm PZK}=\psi,\qquad \psi_{\rm PZK}=\zeta,\qquad
 a B_{\rm PZK}=\chi,\qquad E_{\rm PZK}=-\frac{E}{a^2},
 \label{eq:PZK-metric-dictionary}
\end{equation}
so that their $B$ is precisely the rescaled shear potential $\widetilde\chi=\chi/a$.

In a general gauge, the scalar field is perturbed as $\tilde\phi(t,\vec x) = \phi(t) + \delta \phi(t,\vec x)$. However, when studying scalar perturbations, it is most convenient to work in the unitary gauge (also called uniform scalar field gauge), where $\delta\phi=0$ and the scalar-field fluctuation is encoded in the metric perturbations. In pure Horndeski theory the single scalar mode may be represented by $\zeta$ after the constraints are removed, whereas the Weyl-squared term makes a second scalar combination involving the shear potential $\chi$ dynamical.
In the shear-free Newtonian gauge instead, it is $\chi=0$ and $\delta\phi\neq 0$, with the Newtonian-gauge lapse and spatial curvature perturbations $\psi$ and $\zeta$ in \cref{eq:FLRW-unitary} equal to the Bardeen potentials \cite{Bardeen:1980kt} $\Psi$ and $-\Phi$, respectively.

The Bardeen potentials are related to the perturbations in unitary gauge by\footnote{Note that in our notation $\psi = \phi_{\rm MW}$, $\zeta=-\psi_{\rm MW}$, $\Phi=\Psi_{\rm MW}$, and $\Psi=\Phi_{\rm MW}$ compared to the notation of Malik and Wands (MW) \cite{Malik:2008im}.}
\begin{align}
    \Phi &= -\zeta-H \chi\,, \label{eq:Phi_Bardeen_unitary}\\
    \Psi &= \psi + \dot\chi \,. \label{eq:Psi_Bardeen_unitary}
\end{align}
Further gauge-invariant scalar variables combine the metric and scalar-field perturbations; in particular, we use the uniform-field curvature perturbation $\mathcal R=-\zeta+H\delta\phi/\dot\phi$, so that $\mathcal R=-\zeta$ in unitary gauge, see appendix~\ref{app:gauge-inv-variables}. The Bardeen potentials are particularly useful dimensionless diagnostics because they enter directly in the local tidal tensor, while their gauge invariance by itself should not be confused with observability.

We emphasize that gauge invariance of a perturbation variable is a necessary, but not sufficient, condition for its (in-)stability to be physically meaningful: any gauge-invariant quantity can be rescaled by an arbitrary power of the scale factor while remaining gauge-invariant, and such a rescaling can be used to produce or remove apparent growth at will. This is why in \cref{eq:FLRW-unitary} we adopt the shear potential $\chi$ in the fixed metric convention, rather than, e.g., $\widetilde{\chi}=\chi/a$, and why we regard the geodesic separation normalized by the background scale factor at the same proper time, $Q^{\hat i}=\xi^{\hat i}(\tau)/a(\tau)$ (cf.~appendix~\ref{app:geodesic-deviation}), as a more reliable diagnostic than the growth of some gauge-invariant variable in isolation. The remaining physical question is how these fields enter a specified observable, including its gradients, time integrals, source and observer terms, and clock prescription. We therefore distinguish growth of the fields from the response of that observable; neither can be dismissed on the basis of a coordinate choice.

\subsection{Classical and phenomenological viability criteria}\label{sec:viability-criteria}
A background solution should not be regarded as physically acceptable merely because it solves the background field equations. There is instead a logical hierarchy of questions that has to be addressed. 
Our calculation concerns the quadratic action and the classical linear solutions of the local theory. There, one has to separately check for the presence of ghost modes, gradient instabilities and growing long-wavelength branches. The signs of the kinetic terms are kept explicit throughout; positivity in one perturbation sector does not establish positivity of the full theory. For a valid EFT interpretation we additionally require the omitted operators and the retained frequencies and curvatures to remain within the regime specified in \cref{sec:Horndeski-Weyl-theory}, schematically
\begin{equation}\label{eq:variousconditions}
    \frac{|R|}{\Lambda_{\rm EFT}^2}\ll1,\qquad
    \frac{|R_{\mu\nu}R^{\mu\nu}|}{\Lambda_{\rm EFT}^4}\ll1,\qquad
    \frac{|R_{\mu\nu\rho\sigma}R^{\mu\nu\rho\sigma}|}{\Lambda_{\rm EFT}^4}\ll1,
\end{equation}
together with analogous conditions for the scalar-field derivatives.

The Bardeen potentials have a fixed dimensionless metric normalization, and the weak-metric expansion requires $|\Phi|,|\Psi|\ll1$, which is a necessary but not a sufficient condition for perturbative control. Their growth is a gauge-invariant statement at first order, but it need not be shared by a local curvature or tidal response. Conversely, a small linear response does not bound the nonlinear corrections. We therefore reserve the interpretation of scalar growth, the weak-metric validity bound and the quantum-prescription issue for \cref{sec:discussion}.

For reference, the broader requirements for a classically controlled and phenomenologically viable configuration can be summarized by the following checklist:
\begin{enumerate}
    \item[(i)] regular local curvature and tidal observables in the physical region of interest;
    \item[(ii)] parametric validity of the effective field theory;
    \item[(iii)] a controlled linear spectrum, with no fast gradient or tachyonic instability in the admissible physical modes and with any ghost sector explicitly identified and consistently treated;
    \item[(iv)] control of the approximation used to predict the specified observables, including the neglected nonlinear corrections and any proposed continuation beyond the linear treatment;
    \item[(v)] compatibility with the observed primordial power spectrum;
    \item[(vi)] compatibility with bounds on primordial non-Gaussianity;
    \item[(vii)] acceptable late-time growth, lensing, and gravitational slip;
    \item[(viii)] where applicable, a smooth and physically interpretable GR/FLRW limit.
\end{enumerate}
The logical status of these conditions is not identical. Conditions (i)--(iv) test the internal consistency of the background and of its perturbative description within the assumed EFT; a failure at this level cannot be repaired by agreement with later observational constraints. Conditions (v)--(vii) are additional requirements for cosmological phenomenology, while (viii) is a useful diagnostic rather than a strict necessity. In this work, the perturbation analysis identifies the linear spectrum of the local theory, before any projection of the ghost sector, relevant to (iii), while (i) is probed through selected first-order curvature and geodesic-deviation diagnostics discussed below. For (iv), we obtain the necessary weak-metric validity bound $|\Phi|,|\Psi|\ll1$ and the resulting duration estimate \cref{eq:metric-duration-bound}; full nonlinear control, which would require going to higher orders in perturbation theory, is not established. Likewise, we do not provide a consistent ghost prescription completing (iii), and we assume the validity of the EFT (ii) rather than proving it. We do not perform the dedicated primordial-spectrum, bispectrum, or late-time large-scale-structure calculations, which would require a specified physical dynamics and initial state, needed for (v)--(vii). Consequently, our results do not rely on a fit to cosmological data, and the growing scalar branch found within the backgrounds studied is interpreted, in light of the criteria above, in \cref{sec:discussion}.

\subsection{Ghosts, tachyonic, and gradient instabilities}
Let us briefly clarify nomenclature and state general conditions for the stability of a field theory, as well as the different potential instabilities that can arise. Following \cite{Sbisa:2014pzo}, consider a scalar perturbation $\varphi$ with quadratic Lagrangian
\begin{equation}
    \calL_\phi = \frac{\epsilon_1}{2} \dot{\varphi}^2 - \frac{\epsilon_2}{2a^2} (\partial_i \varphi)^2 - \frac{\epsilon_3}{2} m^2\varphi^2 \,.
\end{equation}
Here $m^2\geq0$ is taken real, while the coefficients $\epsilon_i$ encode the relative signs and normalizations. Relative to the other healthy sectors of the theory, $\epsilon_1<0$ signals a ghost. A gradient (or Laplacian) instability is present when the propagation speed $c_s^2=\epsilon_2/\epsilon_1$ is negative, and is therefore most dangerous in the large-$k$ regime. A tachyonic instability occurs when the effective mass squared $m_{\rm eff}^2=\epsilon_3m^2/\epsilon_1$ is negative and can lead to growing long-wavelength modes. Thus, after fixing the overall sign relative to the rest of the coupled theory, the healthy conditions are
\begin{equation}
    \epsilon_1>0,\qquad \frac{\epsilon_2}{\epsilon_1}>0,\qquad \frac{\epsilon_3m^2}{\epsilon_1}\geq0\,.
\end{equation}
An overall sign multiplying a completely decoupled action has no classical content by itself; once the field is embedded in a theory containing other sectors, however, the relative sign of its kinetic term is physical and determines whether the mode is a ghost.

In the following we will examine our model for ghost, gradient and tachyonic instabilities. Our main goal is to find a region of parameter space where classical gradient and tachyonic instabilities are avoided, while keeping any ghost modes explicitly identified for a possible treatment by the fakeon prescription (or similar).
We emphasize again that, in this work, ``classically stable'' refers exclusively to the absence
of rapidly growing solutions at the level of linear perturbation
theory. In particular, a ghost degree of freedom may still be classically
stable in this restricted sense if its dispersion relation has real
frequencies. The presence of a negative kinetic eigenvalue nevertheless
implies that the Hamiltonian is not bounded from below and may lead to
severe instabilities once interactions or quantum effects are taken into
account. We therefore distinguish linear classical stability from ghost
freedom and from the full viability of the theory. Classical linear stability of the local equations is a natural first diagnostic for any such treatment, although a prescription that restricts the admissible physical data, such as the fakeon prescription, can also change which branches are physical (cf.~\cref{sec:Introduction}).
\medskip

As is well known \cite{Malik:2008im,Baumann:2022mni}, at the lowest order in perturbation theory scalars, vectors, and tensors do not couple. Thus, we insert the perturbed metric \cref{eq:FLRW-unitary} into \cref{eq:Action-Horndeski-Weyl}, expand the action to second order in the perturbations and study the vector, tensor and scalar sectors separately.
When writing down the resulting Lagrangian densities, we will already perform a spatial Fourier transformation, effectively replacing ${\varphi \partial_i ^2 \varphi \rightarrow -k^2 \varphi^2}$ and $(\partial_i \varphi)^2 \rightarrow k^2 \varphi^2$, where $\varphi$ stands for any perturbation variable and $k$ denotes the wavenumber of a given Fourier mode.

\section{Vector perturbations}\label{sec:vector-perturbations}
Let us begin by considering the second-order action for the vector perturbation $B_i$.
After performing a partial integration and using the background equation \labelcref{eq:Bkg_EoM_00} to eliminate $G_{2}$, we obtain the Lagrangian density
\begin{align}\label{eq:action_vectors}
    \calL_v^{(2)} = & -a^3 \left[ \alpha \left(\frac{k^2}{a^2} \dot{B}_{i}^2 - \frac{k^4}{a^4}B_{i}^2\right) -\frac{\calGH}{4} \frac{k^2}{a^2} B_{i}^2 \right] \,.
\end{align}
The coupling $\alpha$ multiplying the kinetic term $\dot B_i^2$ reveals that the Weyl-term introduces propagating vector degrees of freedom, which are absent in pure GR and Horndeski theory, and in the limit $\alpha\to 0$ the vectors become non-dynamical.
In \cref{eq:action_vectors} we defined the (generally time-dependent) function
\begin{align}
    \calGH \coloneqq \Mpl^2+2 \left[ G_4 -2 X G_{4,X} - X \left( H \dot\phi G_{5,X} - G_{5,\phi}\right)\right] \,,
\end{align}
collecting the effect of the free Horndeski coupling functions. This important combination will also appear in the tensor and scalar sectors. Notably, for both the shift-symmetric and power-law cases, $\calGH$ is a constant, $\dot{\calG}_{\rm H}=0$.
In the two cases, the constant $\calGH$ respectively evaluates to
\begin{align}
    \calGH=
    \begin{cases}
    \Mpl^2+2 \left[G_4-2 XG_{4,X}- XH_I \dot\phi  G_{5,X}\right] \qquad &\text{(shift-symmetric)} \,, \\
    \Mpl^2 + 2 \left[a_4 - 2 X a_{4,X} -X \left(2 p X a_{5,X} -a_5\right)\right] \qquad &\text{(power-law case)} \,.
    \end{cases}
\end{align}

To remove the momentum-dependent prefactor of the kinetic term, we introduce the field $U_i = \sqrt{2} k B_i$ such that the Lagrangian becomes
\begin{align}\label{eq:action_vectors_U}
    \calL_u^{(2)} = & - \alpha a \left[ \frac{1}{2} \dot{U}_{i}^2 - \frac{1}{2}\left(\calM^2+\frac{k^2}{a^2}\right) U_{i}^2 \right] \,,
\end{align}
with the mass $\calM$ given by
\begin{align}\label{eq:M2-definition}
    \calM^2 &\coloneqq \frac{\calGH}{4 \alpha} \,,
\end{align}
showing that in the limit $\alpha\to 0$ the vector mode becomes infinitely heavy and ceases to be a propagating degree of freedom of the theory.
The parameter $\calM^2$ plays a decisive role, as it will resurface in the tensor and scalar sectors where it determines the behavior of the solutions.
The mass of the (potentially ghostly) vector mode introduced by the Weyl term therefore sets a relevant dynamical timescale, against which the Hubble and physical-momentum scales have to be compared. It should not, however, be identified in general with the EFT cutoff $\Lambda_{\rm EFT}$ introduced above. In the minimally coupled, weak-curvature limit, $\calM$ is the mass scale associated with the extra massive spin-2 pole generated by the Weyl-squared operator; its helicity components account for the additional tensor, vector, and scalar polarizations, which explains why the same scale reappears across the perturbation sectors. On a time-dependent Horndeski background these components mix with the scalar-tensor sector and their effective masses need not coincide exactly, so we use $\calM$ primarily as the common Weyl-induced scale defined by \cref{eq:M2-definition}. Whether this scale lies below or above $\Lambda_{\rm EFT}$ is a separate, model-dependent question.

From \cref{eq:action_vectors,eq:action_vectors_U} one can directly read off that for $\alpha>0$ the vector modes are ghosts, because in this case their kinetic term is negative. Thus, to avoid vector ghosts we would need $\alpha < 0$.
To avoid gradient instabilities, one requires the speed of propagation to be positive.
We find that vector modes always propagate luminally,
\begin{equation} \label{eq:sound_speed_vectors}
    c_u^2=1 \,,
\end{equation}
implying that no gradient instability is present.
Demanding that the effective mass squared of the modes remain positive to avoid a tachyonic instability, that is $\calM^2>0$, implies that $\calGH/\alpha>0$. Thus, $\alpha$ and $\calGH$ must share the same sign; either $\alpha>0$ and $\calGH>0$, or $\alpha<0$ and $\calGH<0$.

Assuming $\calGH>0$ (which is the case in the limit $G_4=G_5=0$) we have to demand $\alpha>0$ to avoid classical instabilities, at the cost of introducing a ghost mode in the vector sector. In the opposite case, $\calGH<0$ (which requires a sufficiently large non-minimal contribution from $G_4$ and/or $G_5$), one can avoid tachyonic instabilities, as well as ghost modes, by demanding $\alpha<0$.
\medskip

In summary, by an appropriate choice of the Horndeski functions $G_i$ and the sign of the Weyl-coupling, the vector sector can be made free from ghost, gradient and tachyonic instabilities. This does not remove the tensor ghost discussed in the next section.

\section{Tensor perturbations}\label{sec:tensor-perturbations}
Let us now turn our attention to the tensor sector.
After partial integrations on the terms with mixed orders of time derivatives, and using the background equations \labelcref{eq:Bkg_EoM_ij} to replace $G_{3,\phi}$ and \labelcref{eq:Bkg_EoM_00} to eliminate $G_2$, the Lagrangian for tensor perturbations is given by
\begin{align}\label{eq:action_tensors}
    \calL_t^{(2)} = & \frac{a^3}{4} \left[ 2\alpha\left( -\ddot{h}_{ij}^2 + \left[2\frac{ k^2}{a^2} + \left( 2 H^2+\dot{H}\right) \right] \dot{h}_{ij}^2 - \frac{k^4}{a^4} h_{ij}^2\right)
    +\frac{\calGH}{2}\dot{h}_{ij}^2 - \frac{\calF_{\rm H}}{2} \frac{k^2}{a^2} h_{ij}^2
    \right]
\end{align}
where we defined
\begin{align}
    \calF_{\rm H} = \Mpl^2+2 \left[G_4 - X\left(\ddot\phi G_{5,X} + G_{5,\phi}\right) \right] \,.
\end{align}
Like $\calGH$, $\calF_{\rm H}$ is time-independent for both the shift-symmetric and power-law cases.
Taking $\alpha=0$ we recover the results of \cite{Kobayashi:2011nu,Kobayashi:2019hrl}, showing that tensor modes propagate with a modified sound speed and without ghost or gradient instabilities for $\calGH, \calF_{\rm H}>0$ in pure Horndeski theory. A non-zero value of $\alpha$, however, introduces higher time-derivatives and additional $k$-dependent terms in the action.

\subsection{No-ghost condition}
In order to reduce the order of time derivatives and check for the appearance of ghost modes due to the Weyl-coupling, we introduce an auxiliary field $\chi_{ij}$ (not to be confused with the scalar shear potential $\chi$ in \cref{eq:FLRW-unitary}) in the Lagrangian via
\begin{align}
    \bar{\calL}_t^{(2)} = & \frac{a^3}{4} \left[ -2 \alpha \ddot{h}_{ij}^2 + \calG_t \dot{h}_{ij}^2 - \calF_t h_{ij}^2 + 2\alpha \left(\ddot{h}_{ij} - \chi_{ij} \right)^2\right] \,,
\end{align}
such that its equation of motion is simply given by
\begin{align}\label{eq:EoM-Lagrange-chi}
    \frac{\delta \bar{\calL}_t^{(2)}}{\delta \chi^{ij}}=0 \quad\Rightarrow\quad \chi_{ij}=\ddot{h}_{ij} \,.
\end{align}
For compactness of the equations we defined the $k$-dependent functions
\begin{align}
    \calG_t & = 2\alpha \left(2\frac{ k^2}{a^2} + \left( 2 H^2+\dot{H}\right) \right)+\frac{\calGH}{2} \,, \nonumber \\
    \calF_t &=  2\alpha \frac{k^4}{a^4} + \frac{\calF_{\rm H}}{2}  \frac{k^2}{a^2}\,.
\end{align}
After partial integration on the mixed $\ddot{h}_{ij}\chi_{ij}$ term, the Lagrangian can be written as
\begin{align}
    \bar{\calL}_t^{(2)}
    & = \frac{a^3}{4} \left[\dot{\Xi}^{\rm T} \mathbf{K} \dot{\Xi}  - \calF_t h_{ij}^2  + 12 H\alpha \dot{h}_{ij} \chi_{ij} + 2\alpha \chi_{ij}^2 \right]
    \label{eq:action-tensors-reduced}
\end{align}
where $\Xi^{\rm T} = (h_{ij},\chi_{ij})$, and we defined the kinetic matrix
\begin{align}
    \mathbf{K}= \begin{pmatrix}
K_{11} & K_{12} \\
K_{12} & K_{22}
\end{pmatrix} = \begin{pmatrix}
\calG_t & 2\alpha \\
2\alpha & 0
\end{pmatrix} \,.
\end{align}

The ghost is most transparently identified directly from the kinetic quadratic form,
\begin{equation}
    \dot{\Xi}^{\rm T}\mathbf K\dot{\Xi}
    =\calG_t\dot h_{ij}^2+4\alpha\dot h_{ij}\dot\chi_{ij}\,.
\end{equation}
For $\calG_t\neq0$ one may complete the square to find
\begin{equation}
    \dot{\Xi}^{\rm T}\mathbf K\dot{\Xi}=\calG_t\left(\dot h_{ij}+\frac{2\alpha}{\calG_t}\dot\chi_{ij}\right)^2
    -\frac{4\alpha^2}{\calG_t}\dot\chi_{ij}^2\,.
\end{equation}
It is evident that the two terms necessarily have opposite signs for every $\alpha\neq0$. Equivalently, the kinetic quadratic form is indefinite because
\begin{equation}
    \det\mathbf K=-4\alpha^2<0\,.
    \label{eq:tensor-ghost-det}
\end{equation}
The negative determinant also covers the case $\calG_t=0$. Therefore one tensor combination per polarization is necessarily a ghost for any nonzero Weyl coupling $(\alpha\neq0)$.
As both $h_{ij}$ and $\chi_{ij}$ are transverse-traceless tensors, each has two degrees of freedom, implying that the theory propagates two healthy and two ghost tensor modes.
For completeness, in appendix~\ref{app:no-ghost-tensors} we discuss the explicit diagonalization of the kinetic term.

The sign of $\calGH$ determines which tensor branch connects continuously to the low-energy Horndeski graviton. Indeed, on the $\calM^2>0$ branch, as $\alpha\to0$ at fixed momentum the additional Weyl-induced mode becomes infinitely heavy, whereas the surviving massless tensor has kinetic coefficient proportional to $\calGH$. Thus $\calGH>0$ gives the standard healthy low-energy graviton and assigns the opposite kinetic sign to the massive Weyl mode. Conversely, on the branch $\calGH<0$ the low-energy Horndeski graviton is itself ghostlike. This observation is important when interpreting the scalar/vector ghost-free branch $\calGH<0$, $\alpha<0$: it is not a globally ghost-free branch and does not approach a healthy GR/Horndeski tensor sector as $\alpha\to0$.
\medskip

Because the mass and gradient terms are also mixed, we investigate the classical stability of the tensor system directly from its equations of motion rather than attempting a full field-space diagonalization.

In the following, it is important to distinguish the horizon hierarchy from the high-frequency hierarchy. By ``subhorizon" we only mean $k/a\gg H$, whereas the ultraviolet (UV) approximation used to extract the propagation speeds requires the stronger condition $k/a\gg\max(H,\calM)$ (and, more generally, dominance over the remaining effective mass scales). We therefore do not use ``UV" and ``subhorizon" interchangeably. Conversely, our superhorizon analysis assumes $k/(aH)\ll1$; when we additionally take $k\to0$, this is the strict long-wavelength limit and should not be confused with a high-frequency expansion involving $\calM$.

\subsection{Ultraviolet regime}\label{sec:tensors-UV-regime}
We obtain the fourth-order equations of motion for the tensor perturbations by varying $\bar{\calL}_t^{(2)}$, as given in \cref{eq:action-tensors-reduced}, with respect to $h^{ij}$, and using \cref{eq:EoM-Lagrange-chi} for $\chi_{ij}$. This yields
\begin{align} \label{eq:EoM_tensors}
    \ddddot{h_{ij}}
    +6H\dddot{h_{ij}} 
    + \left(\frac{\calG_t}{2\alpha} +3\left( 3H^2 +\dot{H}\right)\right)\ddot{h}_{ij}
    +\frac{1}{2\alpha}\left(3H\calG_t+\dot{\calG}_t\right)\dot{h}_{ij}
    +\frac{\calF_t}{2\alpha}h_{ij} =0 \,.
\end{align}
We first check for gradient instabilities in the UV regime, that is $k/a\gg\max(H,\calM)$, where the momentum-dependent terms dominate over the Hubble and mass scales. This limit is independent of the subclass considered since we can neglect any time dependence of the coefficients.

Using the WKB approximation, we may write the solution for the tensor modes as
\begin{align} \label{eq:WKB-tensors}
    h_{ij}(t,k) = \tilde{h}_{ij}(t)e^{-i\int \omega(t,k) \text{d}t} \,,
\end{align}
and take $\dot{h}_{ij} \approx -i\omega h_{ij}$ at leading WKB order. This approximation is valid as long as the oscillation frequency, the mode amplitudes and the background coefficients vary slowly in time, $|\dot{\omega}| \ll \omega^2$. Keeping the leading, fourth-order terms in $k/a$ and $\omega$, we then find
\begin{align}
    \left(\omega^2-\frac{k^2}{a^2}\right)^2 \approx 0 \,,
    \label{eq:tensor-principal-characteristic}
\end{align}
in agreement with the leading result of \cite{DeFelice:2023psw}, showing that the Weyl-term exclusively dictates the small-scale behavior. Tensor perturbations therefore propagate at the speed of light, $c_t^2 \approx 1$, at leading order in the UV regime. Since this characteristic is a repeated root, the subleading terms decide whether the evolution remains bounded. At that order, the time derivatives of the frequency and of the mode amplitudes must be retained as well, so that replacing every time derivative by $-i\omega$ is no longer sufficient.

A simple finite-momentum check is obtained in a weak-curvature interval in which the background can be frozen on the relevant timescales. Defining $\kappa\coloneqq k/a$ and $r_T\coloneqq\calF_{\rm H}/\calGH$, and retaining the Weyl-induced mass in \cref{eq:EoM_tensors}, we find
\begin{align}
    0&\simeq\omega^4-\left(2\kappa^2+\calM^2\right)\omega^2+\kappa^4+r_T\calM^2\kappa^2 \,,\nonumber\\
    \omega_\pm^2&\simeq\kappa^2+\frac{\calM^2}{2}\pm\frac12\sqrt{\calM^4+4\calM^2(1-r_T)\kappa^2} \,.
    \label{eq:tensor-frozen-dispersion}
\end{align}
For $\calM^2>0$, real frequencies at every $\kappa$ require $0\leq r_T\leq1$. If $r_T>1$, the discriminant becomes negative for $\kappa>\calM/(2\sqrt{r_T-1})$, while for $r_T<0$ the lower branch has $\omega_-^2<0$ for $\kappa^2<-r_T\calM^2$. These conditions are relevant when the corresponding momenta lie below the EFT cutoff and the frozen-background approximation applies; otherwise, the expansion has to be retained in \cref{eq:EoM_tensors}. We examine next the large-scale behavior and check for possible tachyonic instabilities.

\subsection{Superhorizon scales}
On superhorizon scales we can drop the terms involving $k$, since $k\ll aH$. Further, in this limit we have $\mathcal{F}_t\simeq0$ and $\mathcal{G}_t \simeq 2\alpha \left( 2 H^2+\dot{H}\right)+\calGH/2$.
Let us discuss the two subclasses presented in \cref{sec:shiftsymmetrydS,sec:powerlawsolutions} separately. In contrast to the UV limit, the explicit time dependence of the coefficient plays a crucial role in the superhorizon limit.

\subsubsection{Shift-symmetric dS case \label{sec:shiftsymmetrystability}}
In the shift-symmetric case, the equation of motion simplifies to
\begin{align}
    \ddddot{h_{ij}} 
    +6H\dddot{h_{ij}} 
    +\left(11H_I^2 + \calM^2 \right)\ddot{h}_{ij} 
    +3H_I\left(2H_I^2+\calM^2\right)\dot{h}_{ij} \approx 0\,.
\end{align}
We find the general solution to this equation as
\begin{align}
    h_{ij} = \left(c_1+c_2 e^{-3N}  + c_{3,4} \exp\left[-\frac{N}{2} \left(3 \pm \sqrt{1-4 \frac{\calM^2}{H_I^2}}\right) \right] \right)\epsilon_{ij}
    \label{eq:tensor_sol_super}
\end{align}
with the polarization tensor $\epsilon_{ij}$ and the number of $e$-folds $N=H_I t$.
This solution always contains one constant and one decaying branch.
The behavior of the solutions $\propto c_{3,4}$ depends on the signs and values of $\alpha$, $\calGH$ (through $\calM^2$, cf.~\cref{eq:M2-definition}) and $H_I$. If $\calGH$ and $\alpha$ have the same sign, the two additional branches decay, with damped oscillations when $4\calM^2>H_I^2$. At $4\calM^2=H_I^2$, they become $e^{-3N/2}$ and $Ne^{-3N/2}$, which are both bounded.
An instability could in principle appear if $\calM^2<0$, but we excluded this possibility already on the basis that it would introduce a tachyonic instability in the vector sector.
For completeness, we state the precise stability condition for the case $\calM^2<0$: Demanding that the ``$-$" branch of the last term in \cref{eq:tensor_sol_super} does not grow exponentially %
leads to the condition $2 H_I^2 >|\calM^2|$, such that the tensor sector alone exhibits stability also for a small tachyonic mass.

\subsubsection{Power-law solutions}\label{sec:tensors-superhorizon-power-law}
For the power-law case, we also consider the equation of motion \cref{eq:EoM_tensors}, and send $k\to0$.
For power-law expansion, the late-time superhorizon limit is an asymptotic physical regime for the accelerating case $p>1$, but for $0<p<1$ a fixed mode eventually re-enters the horizon. In that case the formulas below describe only an interval outside the horizon or an ordered long-wavelength limit, not a fixed-$k$ late-time prediction.

The resulting equation can be solved in terms of regularized, generalized hypergeometric functions to yield
\begin{align}
    h(t) =& \, c_1 +
    c_2 t^{1-p} \, _1\tilde{F}_2\left(\frac{1-p}{2};\frac{p+1}{2},\frac{3-p}{2};-\frac{1}{4} \calM^2 t^2\right) \nonumber \\
    &+c_3 t^{2-2 p} \, _1\tilde{F}_2\left(1-p;\frac{3}{2}-\frac{p}{2},2-p;-\frac{1}{4} \calM^2 t^2\right) \nonumber \\
    &+ c_4 t^{3-3 p} \, _2\tilde{F}_3\left(1,-\frac{3(p-1)}{2};2-p,\frac{3-p}{2},\frac{5-3 p}{2};-\frac{1}{4} \calM^2 t^2\right) \,,
\end{align}
where we separated the polarization tensor from the mode functions as $h_{ij}(t)=h(t)\times\epsilon_{ij}$.
In a slight abuse of notation, here and throughout the paper we reuse the same labels $c_1$ to $c_4$ for any constant prefactors of different solution branches in the respective equations.
For $\calM^2>0$, the late-time branches have the envelopes
\begin{equation}
    h(t)\sim c_1
      +t^{-3p/2}\left(c_2e^{i\calM t}+c_3e^{-i\calM t}\right) +c_4t^{1-3p} \,.
\end{equation}
The fourth branch is logarithmic at $p=1/3$ and grows for $0<p<1/3$. Therefore, bounded tensor evolution on superhorizon scales requires $p>1/3$, which includes the accelerating backgrounds we are interested in, as well as the same condition as found previously in the vector sector, namely that $\calM^2>0$. These long-wavelength statements complement the finite-momentum condition in \cref{eq:tensor-frozen-dispersion}.

\section{Scalar perturbations}\label{sec:scalar-perturbations}
We first present our analysis in the unitary gauge, and discuss the stability of the lapse perturbation $\psi$, the curvature perturbation $\zeta$, and the shear potential $\chi$. Later we reconstruct the Bardeen potentials $\Phi$ and $\Psi$ using \cref{eq:Phi_Bardeen_unitary,eq:Psi_Bardeen_unitary}, finding that at least one of them inherits a growing branch. As a consistency check, we also compute the evolution of $\Phi$ and $\Psi$ and derive the closed, fourth-order equations for the Bardeen potentials directly, thereby confirming the same gauge-invariant mode growth.

Adopting the notation of \cite{Kobayashi:2011nu}, we write the Lagrangian density for the scalar perturbations of the metric \cref{eq:FLRW-unitary} in unitary gauge (with $\delta\phi=0$) as
\begin{align}
    \calL_s^{(2)} =a^3\bigg[ &-3\calGH\dot\zeta^2 + \calF_{\rm H} \frac{k^2}{a^2}\zeta^2 + \Sigma \psi^2 + 2 \Theta \frac{k^2}{a^2} \chi\psi + 6\Theta \psi \dot\zeta - 2\calGH\frac{k^2}{a^2}\chi\dot\zeta + 2\calGH\frac{k^2}{a^2}\psi\zeta \nonumber \\
    &-\alpha \frac{4}{3} \frac{k^4}{a^4}\left(\dot \chi-H\chi + \psi-\zeta\right)^2
    \bigg] \,,
    \label{eq:action_scalars}
\end{align}
where we defined
\begin{align}
    \Sigma =&\, X \left(G_{2,X} + 2 X G_{2,XX}\right) + 6H\dot\phi X\left(2G_{3,X} + X G_{3,XX}\right) - 2X\left(G_{3,\phi} +X G_{3,\phi X}\right) \nonumber \\
    &- 3 H^2 \left(\Mpl^2 + 2 G_4\right) \nonumber \\
    &+6 H^2X\left(7G_{4,X} +16X G_{4,XX} + 4 X^2 G_{4,XXX} \right) - 6 H\dot\phi\left(G_{4,\phi} +5XG_{4,\phi X}+2X^2 G_{4,\phi XX} \right) \nonumber \\
    &+H^3\dot\phi X \left(30G_{5,X} +26X G_{5,XX} + 4X^2 G_{5,XXX} \right) - 6H^2X\left(6G_{5,\phi} +9XG_{5,\phi X}+2X^2G_{5,\phi XX} \right) \,,
\end{align}
and
\begin{align}
    \Theta =& -\dot\phi X G_{3,X} + H \left(\Mpl^2 + 2 G_4\right) - 8HX \left(G_{4,X}+XG_{4,XX}\right) + \dot\phi \left(G_{4,\phi} + 2XG_{4,\phi X}\right) \nonumber \\
    &-H^2\dot\phi X \left(5 G_{5,X} + 2X G_{5,XX}\right) + 2HX\left(3G_{5,\phi}+2XG_{5,\phi X}\right) \,.
\end{align}
Remarkably, the two functions $\calGH$ and $\calF_{\rm H}$ from the tensor sector appear again in the scalar Lagrangian \cref{eq:action_scalars}, and only two further background combinations, $\Sigma$ and $\Theta$, are required to describe the full dynamics of the Horndeski sector. In \cref{eq:action_scalars}, $\Sigma$ multiplies the quadratic lapse term $\psi^2$, while $\Theta$ controls the mixing of the lapse/shift sector with the curvature perturbation.
In the shift-symmetric case, $\Sigma$ and $\Theta$ reduce to constants, while for the power-law solutions their time-dependence takes the simple forms $\Sigma(t)=\Sigma_1 / t^2$ and $\Theta(t)=\Theta_1 / t$, with $\Sigma_1$ and $\Theta_1$ constant.

When written in the form \cref{eq:action_scalars}, the effect of the Weyl-squared term becomes apparent: it introduces fourth-order derivatives in the action (represented by the $\alpha k^4$-terms in Fourier space), as well as a new dynamical degree of freedom due to the kinetic term $\propto \alpha \dot \chi^2$.
Note that the shear potential $\chi$ enters the metric through $\partial_i\chi$, cf.~\cref{eq:FLRW-unitary}, and is therefore not itself a locally normalized perturbation variable in the homogeneous limit. In the scalar sector of the Weyl-squared operator, its kinetic contribution carries a factor $\alpha k^4/a^4$. The vanishing of the coefficient of $\dot{\chi}^{\,2}$ as $k\to0$ does not, by itself, imply strong coupling. Indeed, the corresponding momentum dependence is shared by the gradient sector, so that the propagation speeds of both scalar eigenmodes remain finite rather than becoming singular. This behavior is consistent with using a geometrically normalized variable proportional to $(k^2/a^2)\chi$. A genuine strong-coupling diagnosis would instead require examining the interaction terms after canonical normalization of the propagating eigenmodes. In appendix~\ref{app:two-scalar-normalization} we derive the finite-$k$ kinetic matrix and show that the leading kinetic coefficient of the shear indeed singles out $\mathfrak s=(k^2/a^2)\chi$.

Note that in \cref{eq:action_scalars}, the field $\psi$ appears without a kinetic term, and is therefore non-dynamical (an auxiliary field). Its equation of motion can be solved algebraically to yield
\begin{align} \label{eq:psi-solution}
    \psi =& \,\frac{\left(4 \alpha k^4 + 3k^2 a^2\calGH\right) \zeta + 9a ^4\Theta\dot\zeta + 3 k^2 a^2 \Theta \chi -4\alpha k^4 (\dot \chi - H\chi)}{4\alpha k^4 -3a^4\Sigma} \,.
\end{align}
This requires $4\alpha k^4\neq3a^4\Sigma$, which we assume throughout. Substituting the solution back into \cref{eq:action_scalars} yields the reduced action for the dynamical fields $\zeta$ and $\chi$.
Schematically, the action reads
\begin{equation}
    \mathcal{L}_s^{(2)} = \mathcal{G}_\chi \dot{\chi}^2  +  \mathcal{G}_\zeta \dot{\zeta}^2 + \mathcal{F}_\chi \chi^2  + \mathcal{F}_\zeta \zeta^2+ \mathcal{Y}_1 \dot{\chi}\dot{\zeta} +  \mathcal{Y}_2 \dot{\chi}\zeta +  \mathcal{Y}_3 \chi \zeta \,.
\end{equation}
The expressions for the coefficients are lengthy and not particularly illuminating, so we refrain from showing them explicitly.
After integrating out $\psi$ by using \cref{eq:psi-solution}, we vary the action with respect to $\chi$ and $\zeta$ to obtain their respective equations of motion, which take the form
\begin{align}
    \ddot{\zeta} + \mathcal{C}_{\dot{\zeta}} \dot{\zeta} + \mathcal{C}_\zeta \zeta + \mathcal{C}_{\ddot{\chi}}\ddot{\chi} + \mathcal{C}_{\dot{\chi}}\dot{\chi} + \mathcal{C}_\chi \chi & = 0 \,, \label{eq:Scalars_EoM_Phi} \\
    \ddot{\chi} + \mathcal{D}_{\dot{\chi}}\dot{\chi} + \mathcal{D}_\chi \chi + \mathcal{D}_{\ddot{\zeta}}\ddot{\zeta} + \mathcal{D}_{\dot{\zeta}} \dot{\zeta} + \mathcal{D}_\zeta \zeta &=0 \,, \label{eq:Scalars_EoM_E}
\end{align}
with non-trivial coefficients $\mathcal{C}_i$ and $\mathcal{D}_i$.
This system of coupled second-order equations can be decoupled to yield two closed fourth-order equations for $\chi$ and $\zeta$ as follows.

By taking two time-derivatives of \cref{eq:Scalars_EoM_E} we first get an expression for $\chi^{(4)}$ in terms of $\zeta$ and its derivatives, as well as up to third derivatives of $\chi$.
One time derivative of \cref{eq:Scalars_EoM_E} gives an expression for $\chi^{(3)}$, again in terms of $\zeta$ and its derivatives up to third order, and derivatives of $\chi$ up to second order.
By taking one time derivative of both \cref{eq:Scalars_EoM_E,eq:Scalars_EoM_Phi}, solving each for $\chi^{(3)}$ and equating the two sides, we can solve for an expression for $\ddot{\chi}$. Then, taking this expression for $\ddot{\chi}$ and equating it to the one obtained from solving \cref{eq:Scalars_EoM_E} for $\ddot{\chi}$, we obtain an expression for $\dot{\chi}$. Lastly, we solve both \cref{eq:Scalars_EoM_E,eq:Scalars_EoM_Phi} for $\ddot{\chi}$, equate both sides, and insert the expression obtained for $\dot{\chi}$ before, which allows us to solve for $\chi$ in terms of $\zeta$ and its derivatives. Taking two time derivatives of \cref{eq:Scalars_EoM_Phi} and inserting all the expressions derived above, we obtain a closed fourth-order equation for $\zeta$. A similar procedure after switching $\chi$ and $\zeta$ leads to the equation for $\chi$.
Schematically, we obtain
\begin{align}
    \zeta^{(4)} + \mu_3 \zeta^{(3)} + \mu_2\ddot{\zeta} + \mu_1\dot{\zeta} + \mu_0 \zeta &= 0 \,, \label{eq:EoM_zeta_4th}\\
    \chi^{(4)} + \nu_3 \chi^{(3)} + \nu_2\ddot{\chi} + \nu_1\dot{\chi} + \nu_0 \chi &= 0 \,.
    \label{eq:EoM_chi_4th}
\end{align}
In general, the coefficients are complicated, time- and $k$-dependent functions, and we refrain from showing the full expressions here.
The apparent four integration constants in each separate fourth-order equation are constrained by the original coupled system and must not be counted as eight independent modes.
\medskip

We note that in the unitary gauge one fixes $\delta\phi=0$ and, therefore, the degrees of freedom are carried by $\zeta$ and $\chi$, while $\psi$ is non-dynamical. Instead, in the Newtonian gauge, we fix $\chi=0$, and the scalar perturbation variables are represented by the Bardeen potentials $\Phi$ and $\Psi$. The scalar-field perturbation $\delta\phi_N$ follows from the gauge transformation between the two gauges and is not an independent degree of freedom. We can relate $\delta\phi_N$ with $\chi$, e.g., following Appendix F.1 of \cite{Domenech:2024drm}. We find that the scalar-field perturbation in the Newtonian gauge can be expressed through the unitary gauge variables as
\begin{align}
    \delta\phi_N %
    = \dot{\phi} \chi \,.
\end{align}
As in both the shift-symmetric and the power-law case it is $\dot\phi=\text{const.}$, $\delta\phi_N$ thus inherits the behavior of $\chi$, as computed in unitary gauge.
We conclude that solving for the evolution of $\zeta$ and $\chi$ in unitary gauge, together with the relations \cref{eq:Phi_Bardeen_unitary,eq:Psi_Bardeen_unitary}, is sufficient for studying both gauges.

\subsection{Ultraviolet regime}
Let us first focus on the high-$k$ UV regime and check for ghost and gradient instabilities.
Expanding the coefficients for large $k$, the action (up to second order in $k$) can be written as
\begin{align}
    \mathcal{L}_{s,k\to\infty}^{(2)} \simeq & \, a^3\bigg[-3 \calGH \dot{\zeta}^2 + \Sigma\dot{\chi}^2 -6 \Theta\dot{\zeta}\dot{\chi} -2 \left(\Sigma + 3 H \Theta\right)\zeta\dot{\chi} \nonumber \\
    &\phantom{a^3\bigg[}+\left(m_\zeta^2 + \tilde{\calF}_\zeta \frac{k^2}{a^2}\right)\zeta^2 +\left(m_\chi^2 + \tilde{\calF}_ \chi \frac{k^2}{a^2}\right)\chi^2 +\left(g_{\zeta \chi} + \tilde{\mathcal{Y}} \frac{k^2}{a^2}\right)\zeta \chi \bigg] \,,
\end{align}
where we defined
\begin{align}
    \tilde{\calF}_\zeta= &\, \calF_{\rm H} + 2 \calGH \qquad \text{and} \qquad \tilde{\calF}_ \chi= 3 H \Theta + \dot\Theta\,,\\
    \tilde{\mathcal{Y}} = &\, 2 \left( \Theta + 2 H \calGH +  \dot\calG_{\rm H} \right)\,, \\
    m_\zeta^2 =& \, 3 \calGH \calM^2 + \Sigma- 9 H \Theta - 3 \dot\Theta \,, \\
    m_\chi^2 =& \, \frac{3 \Theta^2}{4 \alpha} + 3 H^2 \Sigma + H \dot\Sigma + \Sigma \frac{\ddot a}{a} \,, \\
    g_{\zeta \chi}=& \,2\left(3 \Theta \calM^2 + H \Sigma - 6 H^2 \Theta - 3 H \dot\Theta - 3 \Theta \frac{\ddot a}{a}\right) \,.
\end{align}
Note that $\alpha$ enters the large-$k$ action only through the mass terms $m_\zeta^2$, $m_\chi^2$ and the coupling $g_{\zeta \chi}$.
In terms of the vector of perturbations $\vec\Xi^{\rm T}\coloneqq(\zeta,\chi)$, the kinetic and gradient parts of the action can be written as
\begin{align}
    \mathcal{L}_{s, \rm kin} %
    =\frac{1}{2}\vec{\dot{\Xi}}^T \mathbf{K} \vec{\dot{\Xi}} \,,
    \qquad \text{and} \qquad
    \mathcal{L}_{s, \rm grad} %
    = -\frac{1}{2}\frac{k^2}{a^2}\vec{\Xi}^T \mathbf{G} \vec{\Xi} \,,
\end{align}
where the kinetic and gradient matrices are defined as
\begin{align}
    \mathbf{K} =
    \begin{pmatrix}
        -6 \calGH & -6 \Theta \\
        -6 \Theta & 2\Sigma
    \end{pmatrix}
    \qquad \text{and} \qquad
    \mathbf{G} = -
    \begin{pmatrix}
        2  \tilde{\calF}_\zeta & \tilde{\mathcal{Y}} \\
        \tilde{\mathcal{Y}} & 2  \tilde{\calF}_\chi
    \end{pmatrix} \,,
\end{align}
respectively.

The no-ghost conditions are $K_{11}>0$ and $\det\mathbf{K}>0$, yielding explicitly
\begin{align}\label{eq:no-ghost-scalar}
    \calGH<0 \quad \text{and} \quad 3 \Theta^2 + \calGH \Sigma < 0 \,.
\end{align}
Interestingly, in the UV limit these conditions are independent of $\alpha$, and the Horndeski functions alone dictate the presence (or absence) of a ghost in the scalar sector. The inequalities imply $\Sigma>0$. Together with $\calM^2>0$, they select $\alpha<0$ as the scalar/vector ghost-free branch. On this branch, the full finite-$k$ scalar kinetic matrix is positive definite as well (appendix~\ref{app:two-scalar-normalization}), whereas the low-energy tensor mode is ghostlike, as discussed in \cref{sec:tensor-perturbations}. Note that $\calGH<0$ requires a sufficiently large contribution from $G_4$ and/or $G_5$, since $\calGH=\Mpl^2>0$ when $G_4=G_5=0$.

We now make a WKB ansatz for the perturbation vector, $\vec{\Xi}\sim e^{-i\int\omega(t)\,\text{d}t+i\mathbf{k}\cdot\mathbf{x}}$ with slowly varying amplitudes, such that time-derivatives of $\vec{\Xi}$ yield a factor of the frequency $(-i)\omega$.
Assuming further a dispersion relation of the form $\omega \simeq c_s \frac{k}{a}$ and working in the UV regime, $k/a\gg\max(H,\calM)$ and $4|\alpha| k^4\gg3a^4|\Sigma|$, the momentum and frequency terms dominate. Therefore, for the stability analysis we can drop the mass terms ($\sim m_\zeta^2 ,\, m_\chi^2 ,\, g_{\zeta \chi}$) and the single time-derivative terms ($\sim \zeta \dot{\chi} ,\, \dot{\mathbf{K}} \vec{\dot{\Xi}}$), to find the approximate equation of motion
\begin{align}
    \left(\omega^2 \mathbf{K} - \frac{k^2}{a^2}\mathbf{G}\right)\vec{\Xi} \approx 0 \,.
\end{align}
In order to allow for propagating modes (other than the trivial $\vec{\Xi}=0$ solution), the matrix multiplying $\vec{\Xi}$ must not be invertible, such that we get the condition
\begin{align}
    \det (c_s^2 \mathbf{K} - \mathbf{G}) \approx 0 \,.
\end{align}
This yields a quadratic equation for the sound speed $c_s^2$, with the following solutions
\begin{align}
    c_{s,\pm}^2 = & \, \frac{1}{6 \left(3 \Theta^2 + \calGH \Sigma\right)} \Big(\tilde{\mathcal{F}}_\zeta \Sigma-3 \tilde{\calF}_\chi \calGH + 3 \tilde{\mathcal{Y}} \Theta \nonumber \\
 &\pm \sqrt{
 9 \tilde{\calF}_\chi^2 \calGH^2 + 
  3 \calGH \left(-6 \tilde{\calF}_\chi \tilde{\mathcal{Y}} \Theta + \left(2 \tilde{\calF}_\chi \tilde{\mathcal{F}}_\zeta - \tilde{\mathcal{Y}}^2\right) \Sigma\right) + 
  \tilde{\mathcal{F}}_\zeta \left(36 \tilde{\calF}_\chi \Theta^2 + 
     6 \tilde{\mathcal{Y}} \Theta \Sigma + 
     \tilde{\mathcal{F}}_\zeta \Sigma^2\right)}\Big) \,,
\end{align}
corresponding to two different (both generally time-dependent) propagation speeds for the scalar modes, which have to be real and positive to avoid gradient instabilities. The condition $c_{s,\pm}^2>0$ yields another two constraints on the free Horndeski functions.

Combining the scalar no-ghost conditions \cref{eq:no-ghost-scalar} with the ghost-free vector branch $\alpha<0$, $\calGH<0$ and with $c_{s,\pm}^2>0$ restricts the viable parameter region further. Since $\mathbf K$ is positive definite on this branch, gradient stability is equivalent to positivity of $\mathbf G$. Its $22$ entry requires $3H\Theta+\dot\Theta<0$, i.e.~$\Theta<0$ in the shift-symmetric case and $(3p-1)\Theta_1<0$ for the power-law solutions, together with further bounds on the magnitudes of the Horndeski combinations. In the power-law case, if one additionally demands both scalar propagation speeds to be subluminal, $0<c_{s,\pm}^2\leq1$, the ghost-free and gradient-stable parameter region exists only for $p>3$ among the accelerating solutions. To see this, set $g=-\calGH>0$, $x=-\Theta_1/g>0$ and $f=\calF_{\rm H}/\calGH$. Positivity of $\mathbf G$ implies $(f+2)(3p-1)x>(x+2p)^2$, while $\mathbf K-\mathbf G\succeq0$ implies $f\leq1$. Hence $3(3p-1)>(x+2p)^2/x\geq8p$, i.e.~$p>3$.
Superluminal characteristics are not, by themselves, a ghost, gradient, or tachyonic instability of the type tested here. Whether superluminal characteristics imply an obstruction to a standard Lorentz-invariant UV completion is a distinct question \cite{Adams:2006sv}, not the instability question addressed here.

\subsection{Superhorizon scales}
Next, we focus on the large-scale, superhorizon limit. The reduction is performed at $k\neq0$ before expanding in $k/(aH)$; setting $k=0$ in the action first would remove the shear potential from the scalar decomposition. We expand the coefficients $\mu_i$ and $\nu_i$ appearing in \cref{eq:EoM_zeta_4th,eq:EoM_chi_4th} for $k/(aH)\ll1$ and solve the resulting equations of motion for the two cases under consideration separately.

\subsubsection{Shift-symmetric case}
In the superhorizon regime, the coefficients of the fourth-order equations of motion \cref{eq:EoM_zeta_4th,eq:EoM_chi_4th} at leading order reduce to
\begin{align}
    &\mu_0\approx 0  \,, \qquad &&\mu_1\approx 6 H_I^3+3 H_I\calM^2 \,, \\
    &\mu_2\approx 11 H_I^2+\calM^2 \,, \qquad &&\mu_3 \approx 6 H_I \,,
\end{align}
and
\begin{align}
    &\nu_0 \approx 0 \,, \qquad   &
    \nu_1 &\approx  H_I\calM^2 \,,\\
    &\nu_2 \approx \calM^2- H_I^2 \,, \qquad   &
    \nu_3 &\approx 0 \,,
\end{align}
where we also used the $ij$- and scalar-field background equations to eliminate $G_2$ and $G_{2,X}$.
The constant $\calM^2$, corresponding to the mass of the vector modes, is the one defined in \cref{eq:M2-definition}.
Given these coefficients, \cref{eq:EoM_zeta_4th} has the solution
\begin{align}
    \zeta \simeq c_1 + c_2 e^{-3N} + c_{3,4} \exp\left[ -\frac{N}{2} \left(3\pm \sqrt{1-4 \frac{\calM^2}{H_I^2}}\right)\right] \,,
\end{align}
where we introduced the number of $e$-foldings as $N\coloneqq H_I t$.
Interestingly, the solution for $\zeta$ shows the exact same structure as the one found in the tensor sector, cf.~\cref{eq:tensor_sol_super},
\begin{equation}
    h_{ij} \sim \zeta \times \epsilon_{ij}
\end{equation}
up to the (a priori different) prefactors $c_i$ of the distinct solution branches which are determined by the initial conditions.
As a consequence, the conclusion that $\calM^2>0$, and thus $\mathrm{sgn}(\alpha)=\mathrm{sgn}(\calGH)$, is sufficient for stability holds true also for $\zeta$. Through the relation \cref{eq:calR-definition}, this also implies that the uniform-field curvature perturbation $\mathcal R=-\zeta$ is stable in this parameter regime.

At leading order in gradients, the solution \cref{eq:psi-solution} for the auxiliary field $\psi$ reduces to
\begin{align} \label{eq:psi-sol-super}
    \psi \simeq -\frac{\Theta}{\Sigma}\left(3\dot\zeta+\mathfrak s\right) \,,
    \qquad \mathfrak s\coloneqq \frac{k^2}{a^2}\chi \,.
\end{align}
The shear term has to be kept, since on the Weyl-induced branch it is of the same order as $3\dot\zeta$; the strict $k\to0$ expression $\psi\simeq-3\Theta\dot\zeta/\Sigma$ is not uniform on that branch. Using the complete leading-order two-field system derived in appendix~\ref{app:two-scalar-normalization}, the leading lapse vanishes on the massive branches and decays with the decaying branch of $\zeta$, so $\psi$ introduces no additional instability.

For the shear potential $\chi$ we obtain the solution\footnote{Note that the original coupled equations enforce relations between the coefficients $c_i$ and $d_i$ for every admissible solution, leaving only four independent integration constants. At $4\calM^2=H_I^2$, the massive shear solutions become $e^{N/2}$ and $Ne^{N/2}$.}
\begin{align}\label{eq:chi-scaling-shift}
    \chi \simeq d_1 + d_2 e^{-N} + d_{3,4} \, \exp\left[\frac{N}{2}\left(1\pm  \sqrt{1-4 \frac{\calM^2}{H_I^2}}\right)\right] \,.
\end{align}
Notably, \cref{eq:chi-scaling-shift} contains at least one exponentially growing branch. Geometrically, the shear tensor $\sigma_{ij}$ itself (cf.~\cref{eq:shear-tensor}) contains spatial derivatives of $\chi$ and is therefore gradient-suppressed as $k\to0$.
Looking at the action \cref{eq:action_scalars}, one further finds that the kinetic term of the field $\chi$ is multiplied by a factor of $k^4$, indicating that in the strict $k\to 0 $ limit the field becomes non-dynamical, which raises the question whether this superhorizon growth could reflect a gauge or constraint artifact, rather than a physical instability. However, the coefficient of $\dot\chi^2$ degenerates only in the exactly homogeneous limit: at any finite long wavelength, the growing branch is a genuine solution of the coupled system, as the finite-$k$ reduction in appendix~\ref{app:two-scalar-normalization} shows.
These facts make it essential to distinguish growth of the potential $\chi$ from a local physical instability. We therefore next examine gauge-invariant metric variables and, subsequently, local curvature and tidal observables.
\medskip

Using the above solutions for the scalar metric perturbations in unitary gauge, we can compute the gauge-invariant Bardeen potential $\Phi$ as defined by \cref{eq:Phi_Bardeen_unitary}. Importantly, we find that through the term $H_I \chi$, $\Phi$ inherits the growing branch of $\chi$.
Furthermore, the second Bardeen potential $\Psi$, cf.~\cref{eq:Psi_Bardeen_unitary}, receives a contribution $\dot\chi$, which inherits the same growth because of the exponential form of the solution \cref{eq:chi-scaling-shift}.
We emphasize that, although $\zeta$ and $\chi$ have been computed in a fixed gauge, the corresponding exponential growth of $\Phi$ and $\Psi$ is gauge independent at first order.

The structure of this growth becomes transparent in terms of the variables $\zeta$ and $\mathfrak s=(k^2/a^2)\chi$, whose leading kinetic terms remain regular as $k\to0$ (appendix~\ref{app:two-scalar-normalization}). At leading order in gradients their equations close, and in de Sitter the characteristic equation for solutions $\propto a^\lambda$ reads
\begin{equation}
    \lambda(\lambda+3)\left(\lambda^2+3\lambda+2+\frac{\calM^2}{H_I^2}\right)=0 \,.
    \label{eq:scalar-characteristic}
\end{equation}
For $\calM^2>0$ all nonzero roots have negative real parts, so that both $\zeta$ and $\mathfrak s$ remain bounded for arbitrary initial data of this leading-order system. The Bardeen potentials, on the other hand, are reconstructed as
\begin{equation}
    \Phi=-\zeta-\frac{a^2}{k^2}H\mathfrak s \,, \qquad
    \Psi=\psi+\frac{a^2}{k^2}\left(\dot{\mathfrak s}+2H\mathfrak s\right) \,.
    \label{eq:Bardeen-reconstruction}
\end{equation}
The inverse-gradient factor $a^2/k^2$ turns the decaying massive branches of $\mathfrak s$, which scale as $a^{\lambda_\pm}$ with $\lambda_\pm=-\frac32\pm\frac12\sqrt{1-4\calM^2/H_I^2}$, into the growing branches $\propto a^{\lambda_\pm+2}$ of $\Phi$ and $\Psi$. This factor cannot be removed by a field rescaling without changing the fixed metric normalization of the Bardeen potentials. We discuss the physical interpretation of this growth in \cref{sec:discussion}.

To ensure that our approximations made while computing the solutions for $\zeta$ and $\chi$ do not introduce any spurious growth in the Bardeen potentials, we also study their closed fourth-order equations directly, as explained in appendix~\ref{app:closed-eq-Bardeen}.
Solving in terms of $\Phi$ and $\Psi$ directly becomes computationally more involved, because additional terms appear in the equations when substituting $\zeta=-\Phi - H\chi$ and its derivatives $\dot{\zeta} = -\dot{\Phi} - \dot{H}\chi-H\dot{\chi}$ etc.. Furthermore, the solutions for $\zeta$ and $\chi$ have a simpler analytical structure than the one obtained from solving for $\Phi$ and $\Psi$ directly, and therefore they are more convenient to see the behavior and stability properties of individual modes.

In any case, the solutions we obtain by solving directly the closed fourth order equations reveal the same growing and decaying branches and are fully consistent with the scalings derived from the unitary-gauge fields $\zeta$ and $\chi$.

\subsubsection{Power-law solutions}
The coefficients in the superhorizon limit ($k\to0$) and at late times ($t\to\infty$) are given by
\begin{align}
    \mu_0 & \approx0 \qquad && \mu_1 \approx \frac{3p}{t} \calM^2 \nonumber  \,, \nonumber\\
    \mu_2 & \approx \calM^2 \qquad && \mu_3 \approx  \frac{6p}{t} \,,
\end{align}
yielding the (late-time) solution
\begin{align}
    \zeta(t) \simeq& \, c_1 
    + c_3 \, t^{3-6 p}\, _1\tilde{F}_2\left(1-\frac{3 p}{2};2-3 p,\frac{5}{2}-3 p;-\frac{1}{4} \calM^2 t^2\right) \nonumber \\
    &+c_2 \, _1\tilde{F}_2\left(\frac{3 p}{2}-\frac{1}{2};\frac{1}{2},3 p-\frac{1}{2};-\frac{1}{4} \calM^2 t^2\right)
    +c_4 \,t \, _1\tilde{F}_2\left(\frac{3 p}{2};\frac{3}{2},3 p;-\frac{1}{4} \calM^2 t^2\right) \label{eq:zeta-sol-PL}
\end{align}
Expanding the hypergeometric functions for large argument, we find that the solution branches asymptotically scale as
\begin{align}
    \zeta(t\to\infty) \sim t^0 + t^{1-3p} + t^{-3p/2}  e^{-i \sqrt{\calM^2}t}  \,,
\end{align}
which implies stability of the curvature perturbation for $p>1/3$ (in terms of an effective equation of state, this translates to $w<1$).
Since $\mathcal R=-\zeta$ in unitary gauge, this also implies that the uniform-field curvature perturbation $\mathcal R$ is bounded for $p>1/3$ on superhorizon scales. Looking at the action, we see that $\zeta$ has a kinetic term without $k$-dependence, and thus remains propagating also at large scales $k\to 0$. Its stability is therefore an important feature.

Let us emphasize that the stability analysis in this section applies to all power-law expanding backgrounds regardless of whether they accelerate or decelerate; for $0<p<1$, it describes modes while they are outside the Hubble radius (see \cref{sec:tensors-superhorizon-power-law}). Thus, we demonstrate that the growing scalar branch is not a unique feature of de Sitter space but also applies to power-law expanding cosmological backgrounds, extending beyond the results of previous works \cite{Deruelle:2010kf,Ivanov:2016hcm,Salvio:2017xul,DeFelice:2023psw,Palomares:2026iwt}.
We stress that these vacuum scalar-tensor power-law backgrounds are not equivalent to a matter-supported FLRW solution yielding the same value of $p$. In particular, the stability found for matter domination in Ref.~\cite{Sutton:2025ixz} includes an explicit matter sector, which changes the constraint and perturbation system; a direct comparison therefore requires including the corresponding matter perturbations.

The lapse perturbation $\psi$ follows from \cref{eq:psi-sol-super} together with the leading-order two-field system of appendix~\ref{app:two-scalar-normalization}. Its leading contribution vanishes on the massive branches and otherwise scales as $\psi\propto t^{1-3p}$, which decays for $p>1/3$. Lastly, the coefficients of the equation of motion \cref{eq:EoM_chi_4th} of the shear perturbation reduce to
\begin{align}
    \nu_0 & \approx -\frac{p}{t^2}\calM^2 \qquad && \nu_1 \approx \frac{p}{t} \calM^2 \,, \nonumber \\
    \nu_2 & \approx \calM^2-\frac{(p-2) p}{t^2} \qquad && \nu_3 \approx  0 \,,
\end{align}
which yields the solution
\begin{align}
    \chi(t) \simeq &\, c_1 t -c_2 \, _2F_3\left(-\frac{1}{2},\frac{p}{2};\frac{1}{2},-\frac{\beta}{4}-\frac{1}{4},\frac{\beta}{4}-\frac{1}{4};-\frac{1}{4} \calM^2 t^2\right)
    \nonumber \\
    & +c_{3,4} \, t^{\left(5\pm\beta\right)/2} \, _2F_3\left(\frac{3\pm\beta}{4},\frac{p}{2}+\frac{5\pm\beta}{4};1\pm\frac{\beta}{2},\frac{7\pm\beta}{4},\frac{9\pm\beta}{4};-\frac{1}{4} \calM^2 t^2\right) \,, \label{eq:chi-sol-PL}
\end{align}
where we defined
\begin{align}
    \beta\coloneqq \sqrt{4 p^2-8 p+1}
\end{align}
for compactness.
Taking the asymptotic late-time limit, we find the scaling behavior
\begin{align} \label{eq:chi-scaling-PL}
    \chi(t\to\infty) \sim t + t^{-p} + t^{p/2} e^{-i \sqrt{\calM^2}t} \,,
\end{align}
which reveals at least one growing solution branch for any $p>0$.
As in the shift-symmetric case, we have thus found that the curvature perturbation $\zeta$ is stable (for $p>1/3$), while the shear potential contains a growing mode.
\medskip

After studying the unitary-gauge perturbation variables, we turn our attention to the gauge-invariant Bardeen variables.
Multiplying \cref{eq:chi-scaling-PL} by $H \sim t^{-1}$, we find that the contribution to the Bardeen potential $\Phi$ \eqref{eq:Phi_Bardeen_unitary} scales like
\begin{align}
    H \chi \sim 1 + t^{-1-p} + t^{p/2-1} e^{-i \sqrt{\calM^2}t} \,,
\end{align}
which contains a growing mode for $p>2$. For $1/3<p\leq2$, given that $\calM^2>0$, the branches of $\Phi$ are bounded: a constant branch remains, and at $p=2$ the massive oscillations have a constant envelope.
The derivative term $\dot \chi$ contributes further to the Bardeen potential $\Psi$ \eqref{eq:Psi_Bardeen_unitary} as
\begin{align}
    \dot\chi \sim t^0 + t^{-1-p} + t^{-1+p/2} e^{-i \sqrt{\calM^2}t} + t^{p/2}  e^{-i \sqrt{\calM^2}t} \,.
\end{align}
Here we find that the last term contains an oscillating mode that grows as $t^{p/2}$, inherited from the scaling of \cref{eq:chi-scaling-PL}.
The second Bardeen potential $\Psi$ at late times and large scales thus has a growing branch for all $p>0$.
The growth has appeared again through the shift perturbation, but this time in the derivative term $\dot \chi$.
Note, however, that the growth is slower than the background expansion with $a\sim t^p$.
\medskip

\begin{figure}
  \centering
  \includegraphics[width=0.49\textwidth]{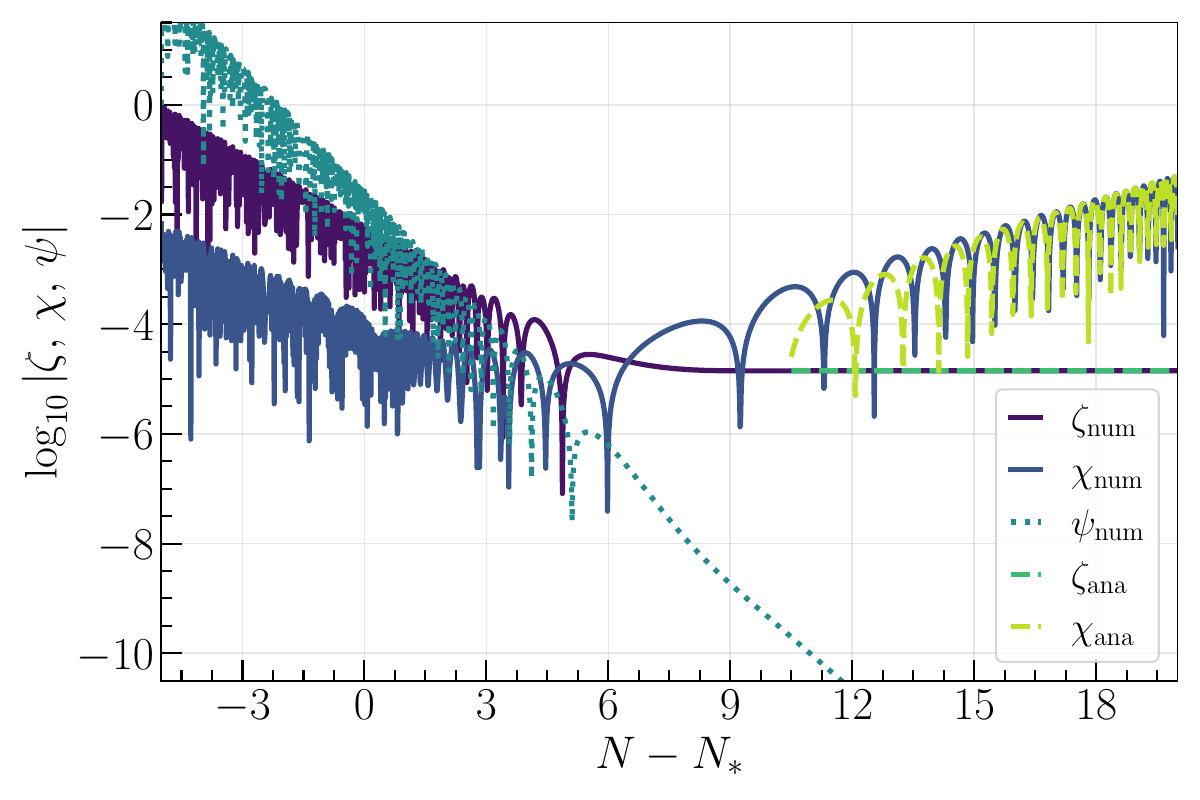}
  \includegraphics[width=0.49\textwidth]{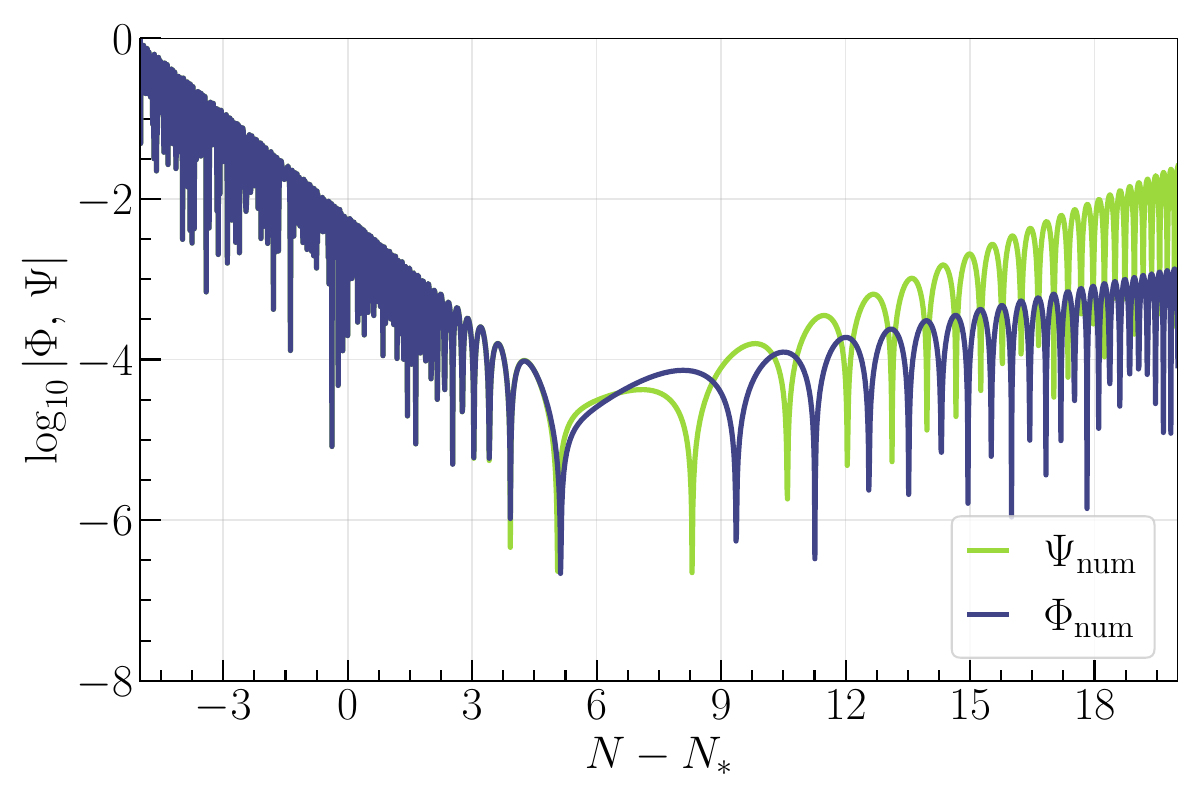}
  \caption{
  Evolution of the metric perturbations $\zeta$, $\chi$ and $\psi$ in unitary gauge (left) and the Bardeen potentials $\Phi$ and $\Psi$ (right) as a function of the number of $e$-folds $N-N_*$ for a set of fixed parameters ($p=4 ,\, \kappa=10^2 ,\, \alpha=\calGH=-1, \,\calF_{\rm H}=-72/77 ,\, \Sigma_1=650 ,\, \Theta_1=-7$), corresponding to the sound speeds $c_{s,1}^2\approx 0.999$ and $c_{s,2}^2\approx 6.63\times 10^{-4}$, and respecting the scalar no-ghost conditions.
  The numerical solutions are plotted together with the analytical superhorizon solutions (dashed curves) for $\zeta$ and $\chi$, which shows the agreement at late times. $N_*$ denotes an arbitrary reference time, and $\kappa=k/k_*$ denotes the ratio of the wavenumber of the mode with respect to the comoving horizon $a_* H_*=k_*$ at that time.
  }
  \label{fig:Pert_Evolution_Plot_1}
\end{figure}

\begin{figure}
  \centering
  \includegraphics[width=0.49\textwidth]{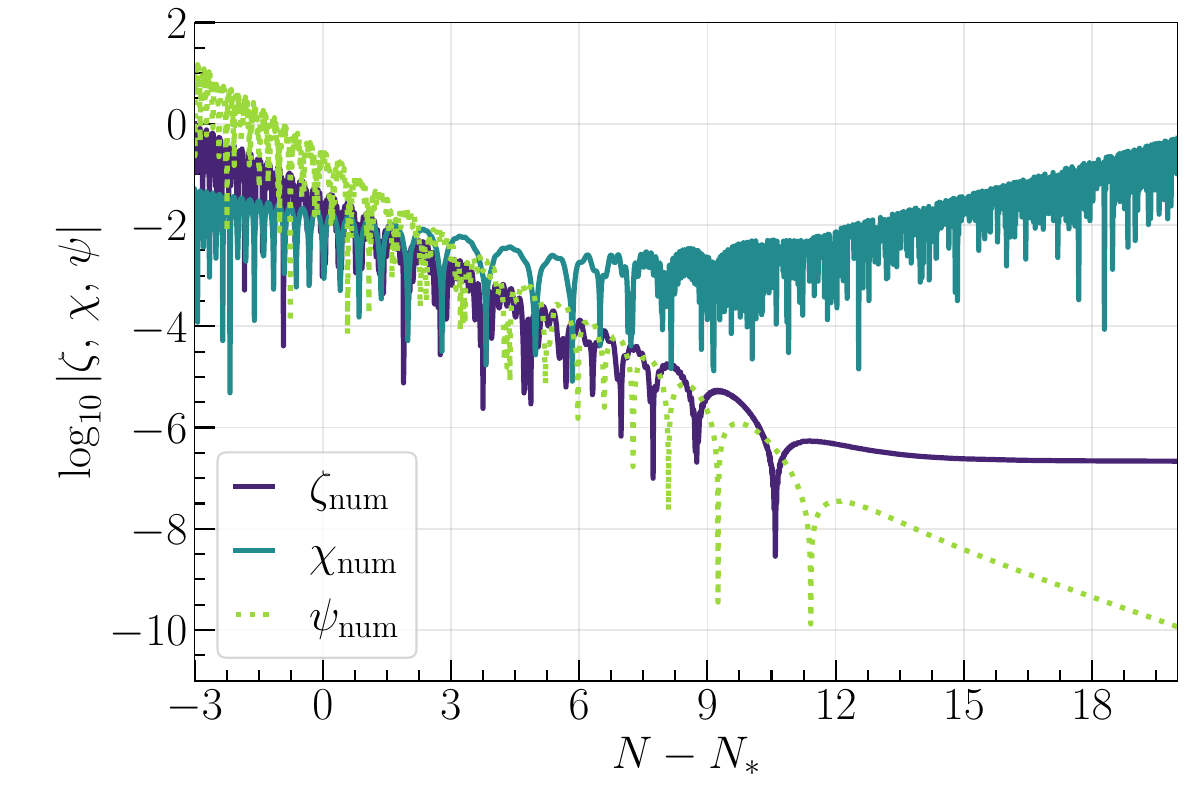}
  \includegraphics[width=0.49\textwidth]{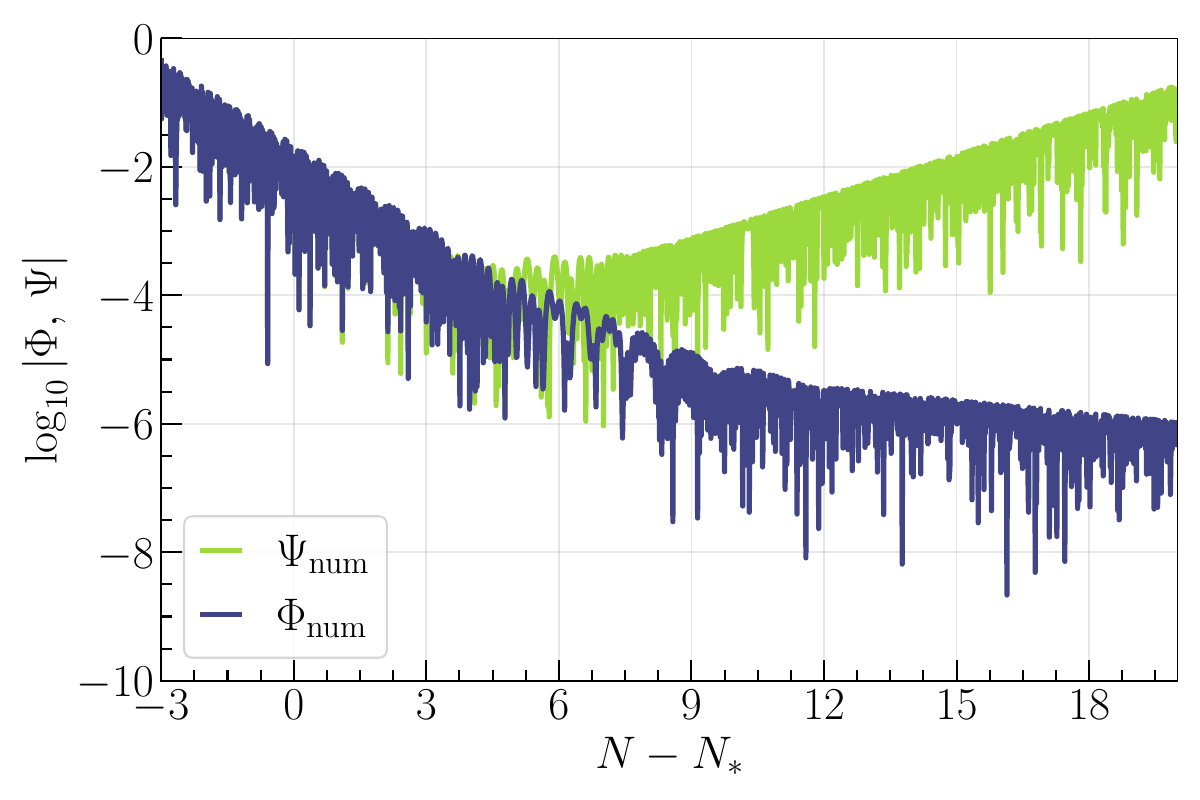}
  \caption{
  Same as \cref{fig:Pert_Evolution_Plot_1}, but for the parameters $p=1.5 ,\, \kappa=10^2 ,\, \alpha=\calGH=-1,\, \calF_{\rm H}=-3,\, \Sigma_1=650 ,\, \Theta_1=-7$, corresponding to the sound speeds $c_{s,1}^2\approx 1.91$ and $c_{s,2}^2\approx 7.78\times 10^{-3}$. This choice of parameters satisfies the no-ghost conditions, but for $p=1.5$ it is not possible to simultaneously enforce both $c_s^2\leq1$. Moreover, $\calF_{\rm H}/\calGH=3$ lies outside the tensor stability range of \cref{eq:tensor-frozen-dispersion}.
  }
  \label{fig:Pert_Evolution_Plot_2}
\end{figure}

\begin{figure}
  \centering
  \includegraphics[width=0.49\textwidth]{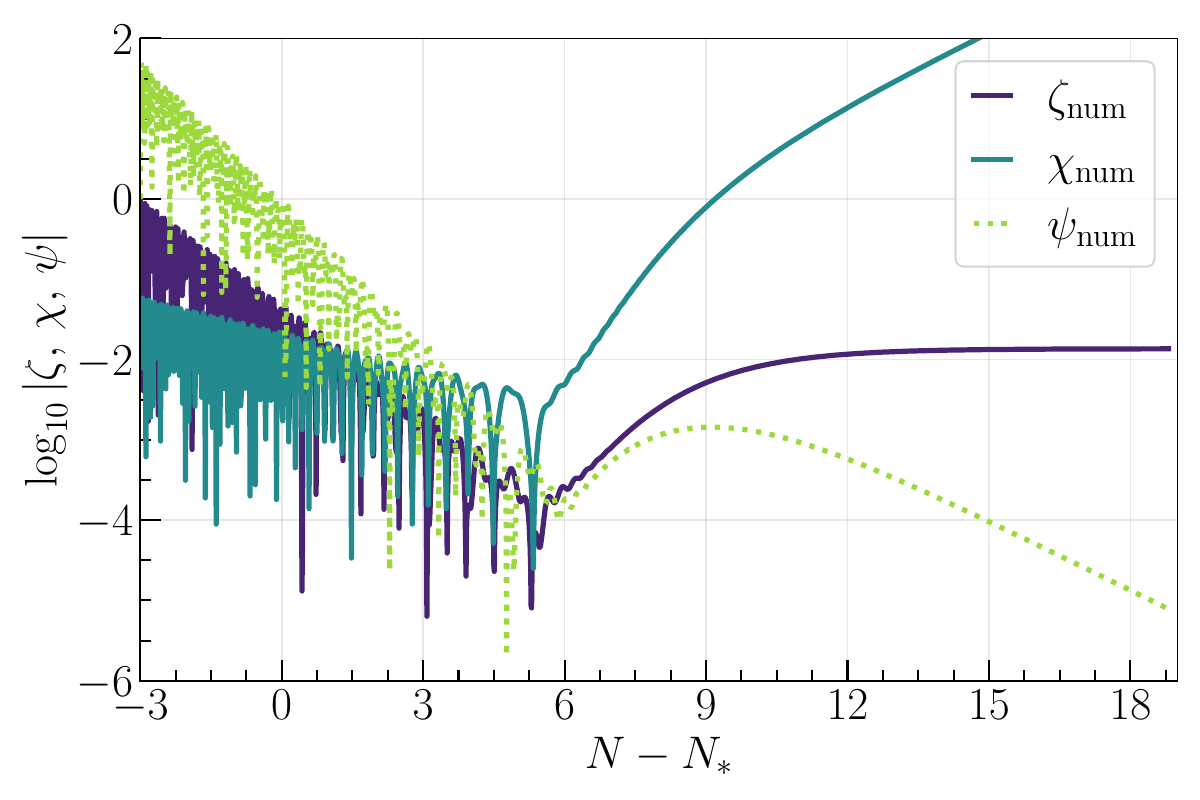}
  \includegraphics[width=0.49\textwidth]{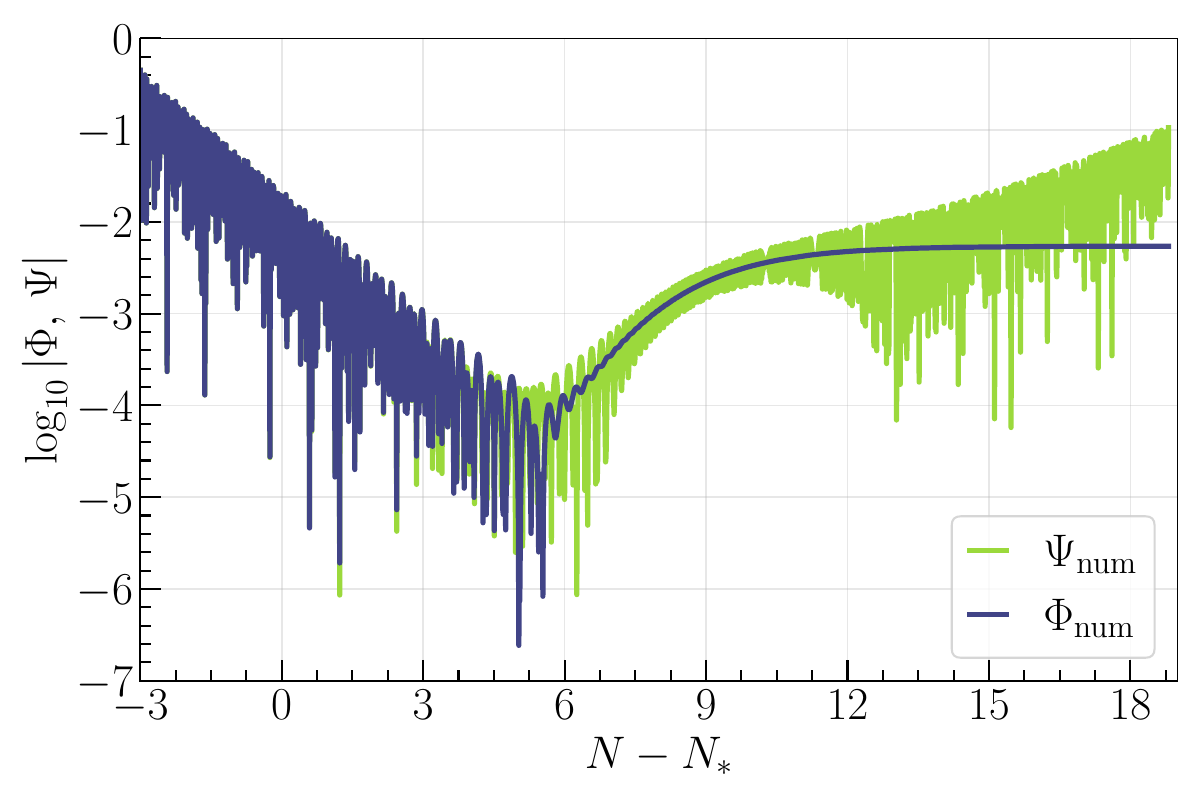}
  \caption{
  Same as \cref{fig:Pert_Evolution_Plot_1}, but for the parameters $p=1.5 ,\, \kappa=10^2 ,\, \alpha=\calGH=\calF_{\rm H}=\Sigma_1=1 ,\, \Theta_1=3$, corresponding to the sound speeds $c_{s,1}^2\approx 0.06$ and $c_{s,2}^2\approx 0.89$. For this choice of parameters both scalar modes propagate sub-luminally, but the no-ghost conditions \cref{eq:no-ghost-scalar} are violated because $\calGH>0$. Note that for these parameter values the oscillations of $\chi$ set in at later times outside the plotted range.
  }
  \label{fig:Pert_Evolution_Plot_3}
\end{figure}

To confirm our analytical approximations, we study the full equations of motion \cref{eq:EoM_zeta_4th,eq:EoM_chi_4th} numerically. In \cref{fig:Pert_Evolution_Plot_1,fig:Pert_Evolution_Plot_2,fig:Pert_Evolution_Plot_3} we show the time evolution of the unitary-gauge perturbation variables, as well as the Bardeen potentials, for some representative parameter values. \Cref{fig:Pert_Evolution_Plot_1} illustrates the excellent agreement of our analytical solutions, \cref{eq:zeta-sol-PL,eq:chi-sol-PL}, with the numerical results at late times. The figures clearly show the stable, oscillating behavior of the perturbations on subhorizon scales, as well as the growth after the respective modes have exited the horizon. The comparison between $p=1.5$ and $p=4$ also shows that, although the number of growing modes changes, in either case at least one perturbation field exhibits a growing superhorizon branch.
%
%
%

\subsection{Curvature invariants}
In order to further clarify the phenomenological viability of our inflating spacetime, we now compute several important scalar curvature invariants and examine their behavior. The linear expressions below are evaluated in unitary gauge and thus represent perturbations on uniform-field slices, i.e.~compared with the background at the same value of the scalar field. In Newtonian gauge, the perturbation of a scalar invariant $I$ with time-dependent background value $\bar I$ would instead have to be corrected by $-\dot{\bar I}\chi$, since $\delta\phi_N/\dot\phi=\chi$; this is the scalar counterpart of the clock matching discussed in appendix~\ref{app:geodesic-deviation}.

Firstly, the linear perturbation of the Ricci scalar is given by
\begin{align}
    \delta R &= 2 \left(\frac{k^2}{a^2}-6 \left(\dot{H}+2 H^2\right)\right)\psi
    +2H \left(2 \frac{k^2}{a^2} \chi-3 \dot\psi\right)
    +2\frac{k^2}{a^2} \dot\chi
    +\frac{4 k^2}{a^2}\zeta+24 H \dot\zeta+6 \ddot\zeta \\
    &\xrightarrow{\, k\to0\,} -12 \left(\dot{H}+2 H^2\right)\psi
    -6H \dot\psi+24 H \dot\zeta+6 \ddot\zeta \,,
\end{align}
while the Kretschmann scalar $\mathcal{K}=R_{\alpha\beta\mu\nu}R^{\alpha\beta\mu\nu}$ at linear order evaluates to
\begin{align}
    \delta \mathcal{K} &= 16H^3 \frac{k^2 }{a^2} \chi
    +8\left(H^2+\dot H\right)\frac{k^2 }{a^2} \left(\dot\chi + \psi\right)-48 \left(2 H^2\dot{H}
    +\dot{H}^2+ 2H^4\right) \psi
    -24 \left(H \dot{H} +H^3\right)\dot\psi\nonumber \\
    &\phantom{=}+16 \frac{k^2 }{a^2}H^2 \zeta + 48 H \left(\dot{H}+2 H^2\right)\dot\zeta+24 \left(\dot{H}+H^2\right) \ddot\zeta \\
    &\xrightarrow{\, k\to0\,} -48 \left(2 H^2\dot{H}
    +\dot{H}^2+ 2H^4\right) \psi
    -24 H\left(\dot{H} +H^2\right)\dot\psi+ 48 H \left(\dot{H}+2 H^2\right)\dot\zeta+24 \left(\dot{H}+H^2\right) \ddot\zeta \,.
\end{align}
As a check, $\bar{\mathcal K}=12[(H^2+\dot H)^2+H^4]$, and a homogeneous constant lapse perturbation with $\zeta=\chi=0$ correctly gives $\delta\mathcal K=-4\psi\bar{\mathcal K}$.
Since the Weyl tensor vanishes on the background, the Weyl-squared invariant has no first-order perturbation. Its leading scalar contribution is quadratic,
\begin{align}
    (C^2)^{(2)} &= \frac{4 k^4}{3 a^4} \left(\psi+\dot\chi-H \chi-\zeta\right)^2 \\
    &\xrightarrow{\, k\to0\,} 0 \,,
\end{align}
and vanishes as $k\to0$ at fixed amplitudes.

In all three cases, the limit $k\to0$ eliminates the explicit terms involving the shear potential $\chi$, leaving behind only combinations involving $\psi$ and $\zeta$. Note, however, that on the massive branches $\dot\zeta=-\mathfrak s/3$ at leading order (appendix~\ref{app:two-scalar-normalization}), so that these terms are of the same order as the explicit shear terms; both decay on accelerating backgrounds. For small, but finite $k$, the curvature invariants receive contributions from $\chi$, which schematically read as
\begin{align}
    \delta R &\sim k^2 \frac{\dot\chi + H \chi}{a^2} \,, \\
    \delta \mathcal{K} &\sim k^2 H^2 \frac{\dot\chi + H \chi}{a^2} \,, \\
    (C^2)^{(2)} &\sim k^4 \left(\frac{\dot\chi + H \chi}{a^2}\right)^2 \,,
\end{align}
where we are only focusing on the scaling with $t$.
In the power-law case, the fastest growing perturbation modes always grow slower than the background, namely $\dot\chi\sim \sqrt{a}$, while in the shift-symmetric case $\chi\sim\dot\chi\sim a^{1/2 + \epsilon_g}$, where the growth exponent $0\leq\epsilon_g<1/2$ (not to be confused with the slow-roll parameter) is determined by the ratio $\calM^2/H_I^2$, cf.~\cref{eq:chi-scaling-shift}. In the limit $\calM^2\ll H_I^2$, the fastest growing mode approaches $\chi\sim\dot\chi\sim a$.
For the accelerating backgrounds considered, the leading shear contributions built from $\left(\dot\chi + H \chi\right)/a^2$ therefore decay at fixed small nonzero $k$. The growth of $\chi$ does not automatically violate the EFT curvature conditions in \cref{eq:variousconditions}.

It is worth distinguishing the local invariant $C^2$ from the conformally invariant density $\sqrt{-g}\,C^2$. For the fastest growing power-law mode one finds $\sqrt{-g}\,C^2\sim(\dot\chi+H\chi)^2/a\sim\mathrm{const.}$, whereas in the shift-symmetric case it scales as $\sqrt{-g}\,C^2\sim a^{2\epsilon_g}$ with $0\leq\epsilon_g<1/2$. The latter growth is not, by itself, a local curvature divergence: $\sqrt{-g}\,C^2$ is a density and contains the expanding physical volume element, while the local scalar $C^2$ continues to decay. We therefore use $C^2$ and the orthonormal-frame curvature components, rather than the integrated density, as diagnostics of local geometric regularity.

\section{Discussion}\label{sec:discussion}
Our linear analysis does not by itself establish either that inflation in this theory is ruled out or that the growing branch is harmless; its main outcome is that the standard linear treatment has a limited domain of validity.
Here, we first delineate the scope and limitations of the current work. Then, we discuss in turn the status of the growing potentials, the physical tidal response, perturbative control, the ghost sector, and the post-inflationary evolution.

\paragraph{Scope and limitations}
Throughout, we study the classical linear perturbations of the local higher-derivative equations derived from \cref{eq:Action-Horndeski-Weyl}, with the Weyl-induced mode retained as a dynamical degree of freedom. By ``classically stable'' we mean the absence of growing linear solutions (see \cref{sec:stability-pert}); this is distinct from ghost freedom. We do not implement a ghost prescription such as the fakeon projection, and we assume, rather than prove, that the relevant scales lie within the domain of validity of the EFT (\cref{sec:Horndeski-Weyl-theory}). Our de Sitter solutions use the exact limit $H=H_I$ with constant $X>0$, and our power-law backgrounds are vacuum scalar-tensor solutions, not matter-supported FLRW solutions with the same expansion rate. For decelerating power laws, $0<p<1$, a mode of fixed $k$ eventually re-enters the Hubble radius; our superhorizon results then describe only the interval during which the mode is outside the Hubble radius, not its late-time evolution. The scalar reduction is performed at $k\neq0$ before expanding in $k/(aH)$, and assumes nondegenerate coefficients, $4\alpha k^4\neq3a^4\Sigma$ and $\calGH\Sigma+3\Theta^2\neq0$. Our ultraviolet results are leading-order WKB statements, and the finite-momentum tensor analysis uses a frozen-background approximation. The numerical solutions use an illustrative initial normalization, not a physical vacuum normalization of the Weyl-induced mode. Finally, we compute selected first-order diagnostics (curvature invariants and tidal fields); a small linear response does not bound nonlinear corrections, and we do not compute primordial spectra, non-Gaussianity or the post-inflationary evolution.
\medskip

\paragraph{The growth of the potentials is genuine.}
The linear growth of the scalar potentials is gauge invariant: the growth of $\Psi$ (and, depending on the background, of $\Phi$) cannot be removed by a change of gauge, and the existence of other variables that remain bounded, such as $\zeta$ or $\mathcal{R}$, does not make the Bardeen potentials bounded.
Gauge invariance is, however, necessary but not sufficient for physical significance, since a gauge-invariant variable can be rescaled by powers of the scale factor, and the shear becomes gradient-suppressed on superhorizon scales. We therefore distinguish the growth of a variable from the response of a specified observable. In this respect, the reconstruction \cref{eq:Bardeen-reconstruction} is instructive: the dynamical fields $(\zeta,\mathfrak s)$ remain bounded in de Sitter, while the Bardeen potentials grow through the inverse-gradient factor $a^2/k^2$ that relates the shear potential to the geometric shear.

At this point, we comment on the discussion of potential gauge artifacts in Ref.~\cite{Palomares:2026iwt}. There, it is argued that the exponentially growing Newtonian-gauge metric perturbations are an artifact because the individual perturbations that they display in flat and comoving gauges remain bounded. Their gauge-invariant analysis, however, contains a useful point of direct comparison with ours. Denoting their gauge-invariant variable called $\chi$ by $\chi_{\rm GI}^{\rm PZK}$ (in order to avoid confusion with our shear potential), we find that their definition reads
\begin{equation}
 \chi_{\rm GI}^{\rm PZK}
 =a^2\left(\frac{B_{\rm PZK}}{a}+\dot E_{\rm PZK}\right)
 +\frac{\delta\phi}{\dot{\bar\phi}} \,.
 \label{eq:PZK-chiGI}
\end{equation}
Using \cref{eq:PZK-metric-dictionary} and imposing only our scalar spatial gauge $E=0$, this becomes $\chi_{\rm GI}^{\rm PZK}=\chi+\delta\phi/\dot{\bar\phi}$. For the Einstein-frame single-scalar system considered in Ref.~\cite{Palomares:2026iwt}, let $\mathcal R$ denote the uniform-field curvature perturbation in our sign convention, $\mathcal R=-\zeta+H\delta\phi/\dot{\bar\phi}$. Together with $\mathcal B=\chi+\zeta/H=-\Phi/H$, one then obtains the exact dictionary
\begin{equation}
 \chi_{\rm GI}^{\rm PZK}
 =\mathcal B+\frac{\mathcal R}{H}
 =\frac{\mathcal R-\Phi}{H} %
 \,.
 \label{eq:PZK-chiGI-dictionary}
\end{equation}
Thus $\chi_{\rm GI}^{\rm PZK}$ is dimensionful, with the natural dimensionless combination $H\chi_{\rm GI}^{\rm PZK}=\mathcal R-\Phi$. In the fully fixed unitary gauge, $\delta\phi=0=E$, \cref{eq:PZK-chiGI} reduces further to the exact identity
\begin{equation}
 \left.\chi_{\rm GI}^{\rm PZK}\right|_{\delta\phi=E=0}=\chi \,.
 \label{eq:PZK-chiGI-unitary}
\end{equation}

Note that Ref.~\cite{Palomares:2026iwt} also finds an exponentially growing superhorizon branch for $\chi_{\rm GI}^{\rm PZK}$ (as well as for the Bardeen potentials) while $\mathcal R$ freezes. In the de Sitter limit, where $H$ is approximately constant, the same branch is therefore present in the dimensionless gauge-invariant quantity $H\chi_{\rm GI}^{\rm PZK}=\mathcal R-\Phi$. This does not by itself establish a physical breakdown -- gauge invariance alone is not a sufficient stability criterion --, but it shows that the growing solution cannot be dismissed solely by noting that a differently normalized metric component is bounded.

There is also a separate residual gauge issue concerning the threading conditions used in Ref.~\cite{Palomares:2026iwt}. In particular, the scalar gauge transformations yield $E_{\rm PZK}\to E_{\rm PZK}+\lambda$ and $B_{\rm PZK}\to B_{\rm PZK}-a\dot\lambda+\lambda^0/a$. Once the temporal slicing has been fixed, either by $\delta\phi=0$ (comoving/unitary slicing) or by $\psi_{\rm PZK}=0$ (flat slicing), the condition $E_{\rm PZK}=0$ fixes the scalar spatial gauge parameter $\lambda$ completely. By contrast, $B_{\rm PZK}=0$ fixes only $\dot\lambda$ and leaves a residual time-independent spatial diffeomorphism $\lambda=\lambda(\mathbf x)$. Thus $\delta\phi=0=B_{\rm PZK}$, or similarly $\psi_{\rm PZK}=0=B_{\rm PZK}$, is an admissible coordinate choice but not a complete gauge fixing without an additional condition. In particular, it should not be interpreted as allowing $E_{\rm PZK}$ and $B_{\rm PZK}$ to be removed simultaneously. Their explicit superhorizon solutions illustrate the distinction: with $E_{\rm PZK}=0$, $B_{\rm PZK}\propto e^{-Ht/2}$ while $aB_{\rm PZK}=\chi\propto e^{Ht/2}$ in unitary gauge; with $B_{\rm PZK}=0$, the same growing branch is carried by $a^2\dot E_{\rm PZK}$.

Our results therefore indicate that the scalar growing branch is present in both Newtonian and fully fixed unitary gauges; whether this growth constitutes a physically unacceptable instability is a distinct question from whether it can be shifted between coordinate variables.
\medskip

\paragraph{Growing potentials do not automatically imply growing tides.}
During inflation the physical separation $\xi^{\hat i}$ of neighboring geodesics grows already in the unperturbed background, so we monitor instead the separation $Q^{\hat i}=\xi^{\hat i}(\tau)/a(\tau)$, normalized by the background scale factor at the same proper time (appendix~\ref{app:geodesic-deviation}); its evolution is governed by the tidal matrix $\mathcal E_{\hat i\hat j}\coloneqq R_{\hat0\hat i\hat0\hat j}$ in
\begin{equation} \label{eq:geodesic-deviation-eq}
    \frac{\text{d}^2\xi^{\hat i}}{\text{d}\tau^2}=-\mathcal E^{\hat i}{}_{\hat j}\xi^{\hat j}\,.
\end{equation}
At linear order, the direct gradient-free contribution of the growing shear to the electric part of the Riemann tensor cancels once the perturbed geometry is compared with the background at the same proper time $\tau$ (cf.~appendix~\ref{app:geodesic-deviation}), and the electric part of the Weyl tensor decays on the inflationary branches studied here. The remaining lapse and gradient contributions to the tidal response are not set to zero by this cancellation, and these results do not imply that all observables remain small at higher orders.
\medskip

\paragraph{Small gradients do not guarantee perturbative control.}
Interaction terms of different order in the perturbations can carry the same gradient suppression. Schematically, for vertices with a common physical derivative scale $\kappa=k/a$ and potential amplitude $A$, $\mathcal L^{(m)}/a^3\sim c_m\Mpl^2\kappa^2A^m$, so that $|\mathcal L^{(m)}|/|\mathcal L^{(2)}|\sim|c_m/c_2|\,|A|^{m-2}$ depends on both the amplitude and the vertex coefficients. Once the growing mode is excited and the corresponding phase lasts long enough, $|\Phi|,|\Psi|\ll1$ ceases to hold and the usual perturbative expansion in Newtonian gauge breaks down. If the envelope $A_N=\max(|\Phi|,|\Psi|)$ grows as $A_N\simeq A_*e^{\sigma\Delta N}$ from an initial amplitude $A_*$, the weak-metric condition $A_N<\varepsilon_{\rm met}\ll1$ restricts the duration of the growing phase to
\begin{equation}
    \Delta N<\frac{1}{\sigma}\ln\frac{\varepsilon_{\rm met}}{A_*} \,,
    \label{eq:metric-duration-bound}
\end{equation}
with $\sigma=1/2$ on the oscillatory power-law branch and $\sigma=1/2+\epsilon_g$ in de Sitter. Note that the primordial adiabatic amplitude does not automatically set $A_*$ for the Weyl-induced mode. A cancellation in a linear observable does not show that the neglected nonlinear terms are small. Since part of the growth resides in a lapse-type potential, it remains to be clarified at second order whether this breakdown reflects a physical failure of perturbation theory or only of the Newtonian-gauge expansion. Gradient-suppressed couplings can also become important at finite $k$: the secondary production of gravitational waves is proportional to gradients squared of the potentials, e.g.~$\partial_i\Psi\partial_j\Psi$ \cite{Domenech:2021ztg}, and the observable gravitational-wave strain is directly related to the transverse-traceless components $h_{ij}$ in Newtonian gauge \cite{Domenech:2025ccu}, also in modified gravity \cite{Domenech:2024drm,Domenech:2025zvi}. This makes $\Phi$ and $\Psi$ well-motivated diagnostic variables, and uncontrolled growth of $\Psi$ could lead to a large backreaction, although the physical strain requires the complete set of source terms and the tensor response. All statements are understood to hold within the domain of validity of the EFT, \cref{eq:variousconditions}. The cutoff $\Lambda_{\rm EFT}$ is conceptually distinct from the Weyl-induced scale $\calM$: if $\calM\gtrsim\Lambda_{\rm EFT}$, the extra pole lies at or beyond the range in which the truncated theory can be trusted, whereas for $\calM<\Lambda_{\rm EFT}$ it can be retained as a dynamical mode, provided the omitted operators and interactions remain controlled. The growth of $\chi$ does not by itself violate \cref{eq:variousconditions} (see the curvature invariants above).
\medskip

\paragraph{Ghost sector.}
The local theory contains an unavoidable tensor ghost whenever the Weyl mode is retained (\cref{sec:tensor-perturbations}). A ghost has a kinetic term of the wrong sign, which is a Hamiltonian and quantum pathology, but its linear classical solutions may remain bounded; ghost freedom and linear classical stability must therefore be distinguished. In particular, the scalar/vector ghost-free branch $\calGH<0$, $\alpha<0$ has a ghostlike low-energy tensor mode (\cref{sec:tensor-perturbations}) and is therefore not globally ghost-free. Physical primordial predictions require a consistent treatment of this sector, for instance through the fakeon prescription \cite{Anselmi:2017ygm,Anselmi:2018kgz,Anselmi:2020lpp}, which restricts the admissible physical data and changes the induced classical response \cite{Anselmi:2018bra,Anselmi:2025uda}, and has not been implemented for the present model. The nonlinear constraints and the complete observable response must then be checked in the resulting physical theory, starting at second order in perturbations. Likewise, observational bounds on the amplitude, tilt and non-Gaussianity of the primordial spectrum \cite{Planck:2018vyg,Planck:2018jri} can only be confronted after identifying which combination of the two propagating scalar modes corresponds to the observed curvature perturbation; this, as well as the gravitational slip $\eta_{\rm slip}=\Phi/\Psi$ and the lensing potential $\Phi+\Psi$, requires dedicated calculations beyond the scope of this paper.
\medskip

\paragraph{Post-inflationary evolution.}
A viable inflationary scenario must also admit a controlled post-inflationary evolution. The persistence of the growing long-wavelength branch on the decelerating power-law backgrounds studied here makes this an important consistency test. We stress that these vacuum scalar-tensor backgrounds are not equivalent to matter-supported FLRW solutions with the same value of $p$ (cf.~\cite{Sutton:2025ixz}). Establishing whether the growing branch obstructs a realistic radiation- or matter-dominated era requires following the relevant finite-wavelength modes, including the supporting matter sector and the transition from inflation. A physical instability that prevents such an evolution would rule out the corresponding cosmological model, irrespective of its behavior during inflation.

\medskip
We therefore conclude neither that inflation is excluded in this theory nor that its viability has been established.

\section{Summary and conclusion}\label{sec:conclusion}
We have investigated the classical stability of linear scalar, vector, and tensor perturbations in Horndeski theory augmented by a quadratic Weyl term, cf.~\cref{eq:Action-Horndeski-Weyl}, on inflating FLRW backgrounds. Motivated by the expectation that the Weyl ghost may be rendered harmless -- either because it is ``soft'' in the sense of \cite{Smilga:2004cy}, or because it can subsequently be removed via the fakeon prescription \cite{Anselmi:2017ygm,Anselmi:2018kgz} -- we asked whether a classically stable inflationary background exists within this theory. To cover a broad range of inflationary scenarios, we studied two subclasses: shift-symmetric models in the de Sitter limit, and power-law solutions $a\sim t^p$ with accelerated expansion for $p>1$.

In the vector sector, \cref{sec:vector-perturbations}, the Weyl term renders the otherwise non-dynamical vector $B_i$ propagating, with a mode whose mass is set by $\calM^2=\calGH/(4\alpha)$, see \cref{eq:M2-definition}. The vector modes propagate luminally, $c_u^2=1$, so no gradient instability arises, and demanding $\calM^2>0$ excludes the tachyonic instability, fixing $\calGH$ and $\alpha$ to have the same sign. For $\calGH>0$ this requires $\alpha>0$ at the cost of a vector ghost, whereas the branch $\calGH<0$, $\alpha<0$ (which requires a sufficiently large contribution from $G_4$ and/or $G_5$, since $\calGH=\Mpl^2>0$ for $G_4=G_5=0$) avoids ghost, gradient, and tachyonic instabilities within the vector sector.

The tensor sector, \cref{sec:tensor-perturbations}, propagates two healthy and two ghost modes: the higher-derivative structure introduced by the Weyl term always renders one of the two transverse-traceless fields a ghost. Nevertheless, in the UV regime the tensor modes propagate at the speed of light, $c_t^2\approx1$, and on superhorizon scales all mode branches tend to constants or decay, provided $\calGH$ and $\alpha$ have equal sign (and $p>1/3$ in the power-law case). At finite momentum, a frozen-background analysis further requires $0\leq\calF_{\rm H}/\calGH\leq1$. Under these conditions, classical stability is preserved in the tensor sector despite the presence of the ghost, in line with the behavior found for Stelle's theory \cite{DeFelice:2023psw}. On the branch $\calGH<0$, however, the low-energy tensor mode is itself ghostlike.

The scalar sector, discussed in \cref{sec:scalar-perturbations}, is the critical one. Working in unitary gauge, we integrated out the non-dynamical lapse $\psi$, cf.~\cref{eq:psi-solution}, and reduced the dynamics to the curvature perturbation $\zeta$ and the shear potential $\chi$. In the UV regime the no-ghost conditions $\calGH<0$ and $3\Theta^2+\calGH\Sigma<0$ are independent of $\alpha$, so that the Horndeski functions alone dictate the presence of a scalar ghost; together with $\calM^2>0$ this again singles out $\alpha<0$ as the scalar/vector ghost-free branch, and the requirement for the two propagation speeds to be positive, $c_{s,\pm}^2>0$ yields two further constraints on the Horndeski functions. On superhorizon scales, however, we find that while $\zeta$ is stable (for $p>1/3$ in the power-law case, and analogously to the tensor solution in the shift-symmetric case), the shear potential $\chi$ always contains a growing mode, as shown in \cref{eq:chi-scaling-shift,eq:chi-scaling-PL}. Because the Bardeen potentials inherit this growth -- $\Phi$ through the combination $H\chi$ and $\Psi$ through $\dot\chi$, cf.~\cref{eq:Phi_Bardeen_unitary,eq:Psi_Bardeen_unitary} -- at least one Bardeen potential grows on super-Hubble scales for every background we consider.
In the shift-symmetric case, both Bardeen potentials generically inherit a growing branch. In the power-law case $\Psi$ grows for every $p>0$, whereas $\Phi$ is bounded for $1/3<p\leq2$. For the inflationary subset $1<p\leq2$, however, this boundedness of $\Phi$ cannot be combined with the ghost-free requirement and subluminal propagation of both scalar modes, since the latter requires $p>3$. We confirmed these behaviors both by solving the unitary-gauge equations and by solving the closed fourth-order equations for $\Phi$ and $\Psi$ directly and numerically, finding excellent agreement with the analytical superhorizon solutions.
In terms of $\zeta$ and $\mathfrak s=(k^2/a^2)\chi$, whose kinetic terms remain regular at long wavelengths, the scalar fields remain bounded in de Sitter and on accelerating power-law backgrounds; the growth of the Bardeen potentials arises from the inverse-gradient factor $a^2/k^2$ in their reconstruction, \cref{eq:Bardeen-reconstruction}.
We thus find that, for every inflationary background within the classes considered, at least one Bardeen potential grows on superhorizon scales, generalizing the scalar growth found in \cite{Deruelle:2010kf, DeFelice:2023psw, Palomares:2026iwt} beyond the minimal scalar-field case and the Starobinsky model. The stability properties of scalar, vector and tensor modes are summarized in \cref{tab:summary}.
\medskip
\begin{table*}[t]
\centering
\begin{tabular}{|c|c|c|c|c|}
\hline
 & & & \multicolumn{2}{c|}{growing branch (superhorizon)} \\
\cline{4-5}
Mode & ghost & gradient (UV) & shift-symmetric & power-law \\
\hline\hline
$U_i$ & \warning \, ($\alpha<0$) & \healthy \, ($c_u^2=1$) & \multicolumn{2}{c|}{\warning \, ($\mathrm{sgn}\,\alpha=\mathrm{sgn}\,\calGH$)} \\
\hline
$h_{ij},\,\chi_{ij}$ & \pathology & \warning \, ($c_t^2\approx1$; $0\leq\calF_{\rm H}/\calGH\leq1$) & \warning \, (same) & \warning \, (same; $p>1/3$) \\
\hline
$\zeta$ & \multirow{2}{*}{\warning ($\calGH<0,\,3\Theta^2+\calGH\Sigma<0$)} & \multirow{2}{*}{\warning ($c_{s,\pm}^2>0;\,\Sigma>0$)}
 & \healthy & \warning \,($p>1/3$) \\
$\chi$ & & & \pathology & \pathology \\
\hline
$\Phi$ & \multirow{2}{*}{\warning \, (same)} & \multirow{2}{*}{\warning \, (same)} & \pathology & \warning ($1/3<p\leq2$) \\
$\Psi$ & & & \pathology & \pathology \\
\hline
\end{tabular}
\caption{Summary of ghost ($K_{11}>0,\,\det\mathbf{K}>0$), gradient ($c_{s,\pm}^2>0$, $k\to\infty$), and superhorizon ($\calM^2>0$, $k\to0$) (in)stability for each perturbation mode in the vector, tensor and scalar sector. For tensors, the gradient entry includes the finite-momentum condition of \cref{eq:tensor-frozen-dispersion} in the frozen-background approximation. The symbols mean
\healthy: no instability of the indicated type in the stated regime, \warning: pathology avoided under the stated conditions, and \pathology: an unavoidable ghost or a growing branch in the stated regime.
As $\Phi$ and $\Psi$ are gauge-invariant combinations of $\zeta$ and $\chi$, cf.~\cref{eq:Phi_Bardeen_unitary,eq:Psi_Bardeen_unitary}, the no-ghost and sound-speed conditions are the same. The non-dynamical lapse perturbation $\psi$ decays for $p>1/3$ in the power-law case and hence throughout the accelerating regime $p>1$. For $0<p<1$, the superhorizon entries describe modes while they are outside the Hubble radius. Requiring both scalar modes to be subluminal, $0<c_{s,\pm}^2\leq1$, further restricts the ghost-free, gradient-stable accelerating power-law solutions to $p>3$.}
\label{tab:summary}
\end{table*}

If classical boundedness of all local modes is required, as for a perturbative treatment of the Weyl-induced mode as a classical degree of freedom, our findings pose a serious challenge to the viability of the theory as a model of inflation at quadratic order in curvature. A ghost prescription that restricts the admissible physical data, such as the fakeon prescription, may change this conclusion and requires a dedicated analysis.
It is interesting to note, however, that the growth of the Bardeen potentials arises from the inverse-gradient factor relating the shear potential to the geometric shear, and that the growing modes of $\Psi$ and $\chi$ always grow slower than the background expansion with $a$ (or at most as fast as the background, $\chi\sim\Psi\sim a$, for $\calM^2/H_I^2\to 0$ in the shift-symmetric case). Furthermore, the leading shear contributions to $R$, $\mathcal{K}$ and $C^2$ decay in the regime considered, the direct contribution of the growing shear to the clock-matched linear tidal field cancels, and the electric part of the Weyl tensor decays. Whether the growth we found constitutes a physical instability therefore hinges on identifying a genuinely observable diagnostic beyond linear order, which we discussed in \cref{sec:viability-criteria,sec:discussion}.

Higher-order operators or a UV completion may modify the perturbation spectrum around FLRW; whether they do so requires a separate calculation beyond the truncated action studied here.

Because the Weyl term introduces additional scalar and tensor degrees of freedom, it is not immediately clear which combination of perturbation variables should be identified with the curvature perturbation constrained by CMB observations. The standard single-field mapping between $\mathcal R$ (or $\zeta$) and the observed amplitude $A_s$ and tilt $n_s$ assumes a single propagating scalar mode, whereas our system contains two, $\zeta$ and $\chi$. We find that $\zeta$ itself is stable on superhorizon scales (for $p>1/3$), which is a necessary condition for a viable power spectrum, but we have not established what observable, if any, is set by the growing $\chi$ sector, nor whether the gravitational slip $\eta_{\rm slip}=\Phi/\Psi$ provides an independent observational handle on the second degree of freedom. We leave a full treatment of this mapping to future work.

Lastly, it is important to emphasize that in our analysis, we have restricted ourselves to linear perturbation theory. Whether the growing modes we find are tamed (or even amplified) once nonlinear effects are taken into account, remains an open question beyond the scope of this work. Nevertheless, the weak-metric bound \cref{eq:metric-duration-bound} limits the duration over which the Newtonian-gauge linear treatment can be trusted, and secondary gravitational wave production, which depends on gradients of the Bardeen potentials, suggests that the scalar growth may also be important at higher orders in perturbation theory.

\begin{acknowledgments}
    We would like to thank Shinji Mukohyama for collaboration on the early stages of the project. We also thank Alexander Ganz and Misao Sasaki for helpful discussions, and Adrian Palomares and Ying-li Zhang for useful correspondence.
    This work is supported by the DFG under the Emmy-Noether program grant no.~DO 2574/1-1, project number 496592360, and in part by JSPS KAKENHI No.~24K00624.
    J.T.~acknowledges support from the Italian Ministero dell’Università e della Ricerca through the FIS 3 project FIS-2024-02546 “PRIMAVERA” (CUP C53C25001070001), as well as from Istituto Nazionale di Fisica Nucleare (INFN) through the Theoretical Astroparticle Physics (TAsP) project.
    For the derivation of perturbation equations we used the \texttt{xPand (xAct)} package \cite{Pitrou:2013hga} for \texttt{Mathematica}.
    The authors acknowledge the use of Claude Sonnet 5.5 and Opus 5.5 (Anthropic), and ChatGPT (OpenAI) to assist in drafting and proof-reading sections of the manuscript and for exploratory algebraic checks; all scientific content, derivations, and conclusions were verified by the authors, who take full responsibility for the final text.
\end{acknowledgments}

\appendix

\section{Second-order action of GR in Newtonian gauge} \label{app:GR_Newton}
As a warm-up, let us consider the second-order action in pure general relativity in Newtonian gauge. We take the Einstein-Hilbert action
\begin{align}
    S_{\text{EH}}[g_{\mu\nu}] = \frac{\Mpl^2}{2}\int \text{d}^4x \sqrt{-g} \, R \,,
\end{align}
and expand it to second order in perturbations, yielding
\begin{align}
    S^{(2)}_{\text{EH}} = \frac{\Mpl^2}{2}\int \text{d}^3x\text{d}t \,
    a^3 \Bigg[& -6\left(\dot{\Phi}^2 -\frac13a^{-2} \left(\partial_i\Phi\right)^2 \right) +\frac14 \left(\dot{h}_{ij}^2 -a^{-2}\left(\partial_k h_{ij}\right)^2\right) \nonumber\\
    & -\left(9 H^2(\Psi+2\Phi)+12 H \dot{\Phi} - 4 a^{-2}  \partial^i\partial_i\Phi\right)\Psi
    \nonumber\\
    & +\left(3 H^2+2 \dot{H}\right) \left(3\Phi^2 -\frac12 h_{ij}^2\right)
    \Bigg] \,,
\end{align}
where we considered only first order metric perturbations.
As one can see, $\Psi$ appears without time derivatives and is therefore a non-dynamical (auxiliary) field whose equation enforces the Hamiltonian constraint.
Variation of the action with respect to $\Psi$ leads to
\begin{align}
    0=9 H^2 (\Psi +\Phi) + 6 H \dot\Phi +2 \frac{k^2}{a^2} \Phi\,,
\end{align}
in Fourier space.
In pure vacuum GR without matter $H^2 = 0 = 3H^2 + 2\dot{H}$. The constraint arising from variation w.r.t.~$\Psi$ thus enforces $\Phi=0$ for $k\neq0$, and the scalar sector becomes completely non-dynamical, and reduces to the Hamiltonian and momentum constraints.
This reflects the well-known fact that there is no scalar graviton, i.e.~no propagating scalar degree of freedom in pure GR. Further, we see that the tensor modes propagate with $c_t^2=1$.

Adding a matter action $S_m$ introduces couplings of the matter energy density and pressure with the gravitational potential and tensors.
One may worry about the wrong sign of the kinetic term of $\Phi$. However, once matter is added to the theory and $\Phi$ becomes dynamical, it behaves as a normal (non-ghost) field. This is however not so straightforward to see.
By adding e.g.~a scalar field or perfect fluid, one can derive the closed action for scalar perturbations. As shown in Appendix D of \cite{Domenech:2024drm} for the case of Horndeski gravity (which reduces to GR plus a canonical scalar field for $G_4=0$, $G_3=G_5=0$ and $G_2= X-V(\phi)$, in our convention with a separate Einstein--Hilbert term), by computing the action in unitary gauge, integrating out the auxiliary fields and performing a gauge transformation to go back to the Newton gauge, one obtains (in Fourier space)
\begin{align}
    \mathcal{L}_s^{(2)} = \Mpl^2\frac{a k^2}{\epsilon H^2} \left(\dot{\Phi}^2 - c_s^2 \frac{k^2}{a^2}\Phi^2 + H\frac{\dot{\epsilon}}{\epsilon}\Phi^2 \right)
\end{align}
with $c_s^2=1$ for the canonical scalar and $\epsilon=-\dot{H}/H^2>0$. In standard cases one has power-law behavior of the scale-factor $a\sim t^p$ such that $H=p/t$ and $\epsilon=1/p=\text{const.}$ and $\dot{\epsilon}=0$. Thus, one recovers the standard action for a (non-ghost) scalar field, showing that GR is indeed a healthy theory.

\section{No-ghost condition in the tensor sector}\label{app:no-ghost-tensors}
The determinant \cref{eq:tensor-ghost-det} already shows that the tensor kinetic form has one negative direction per polarization. Since $h_{ij}$ is dimensionless while $\chi_{ij}=\ddot h_{ij}$ has dimension of mass squared, an explicit diagonalization requires a common normalization of the two fields. We therefore introduce a constant reference mass $\mu_{\rm ref}>0$ and use the dimensionless pair $\mathbf y^{\rm T}=(h,\chi/\mu_{\rm ref}^2)$ for one tensor polarization. The kinetic matrix inside the prefactor $a^3/4$ then reads
\begin{equation}
    \overline{\mathbf K}=
    \begin{pmatrix}
    \calG_t & 2\alpha\mu_{\rm ref}^2\\
    2\alpha\mu_{\rm ref}^2 & 0
    \end{pmatrix} \,,
    \qquad
    \lambda_\pm=\frac12\left(\calG_t\pm\sqrt{\calG_t^2+16\alpha^2\mu_{\rm ref}^4}\right) \,.
    \label{eq:dimensioned-tensor-eigenvalues}
\end{equation}
For real $\alpha\neq0$, one has $\lambda_+>0$ and $\lambda_-<0$, independently of the choice of reference mass and including the case $\calG_t=0$. Therefore, one of the two fields must be a ghost.

At fixed time and momentum, an orthogonal matrix $T(t,k)$ diagonalizes this kinetic form, $T^{\rm T}\overline{\mathbf K}T=\mathbf D_K$. Since $T$ is time dependent, the field transformation $\mathbf y=T\boldsymbol\xi$ also generates additional terms,
\begin{align}
    \dot{\mathbf y}^{\rm T}\overline{\mathbf K}\dot{\mathbf y}
    =\dot{\boldsymbol\xi}^{\rm T}\mathbf D_K\dot{\boldsymbol\xi}
    +2\dot{\boldsymbol\xi}^{\rm T}T^{\rm T}\overline{\mathbf K}\dot T\boldsymbol\xi
    +\boldsymbol\xi^{\rm T}\dot T^{\rm T}\overline{\mathbf K}\dot T\boldsymbol\xi \,,
    \label{eq:time-dependent-tensor-rotation}
\end{align}
which do not change the signs of the kinetic terms but contribute to the single-derivative and mass terms. Moreover, the mass and gradient terms are still not diagonal, and there is mixing among the different modes. We could bring the action to completely diagonal form by performing another rotation in field space, but in general this cannot be done while preserving the diagonal kinetic form. For simplicity, in \cref{sec:tensors-UV-regime} we therefore investigate the system's classical stability by studying the equations of motion.

\section{Geodesic deviation, gauge-invariant tidal fields, and the \textit{e}-fold description}\label{app:geodesic-deviation}
This appendix discusses geodesic deviation for test probes minimally coupled to the metric $g_{\mu\nu}$ of \cref{eq:Action-Horndeski-Weyl}. This specifies the metric and proper time used by the observers; a change of conformal frame must transform the probe coupling as well. We distinguish the local tidal tensor at an event from its perturbation relative to a background evaluated at a specified clock.
The derivation is organized so as to separate coordinate-dependent quantities from the tidal field measured by a freely falling observer.
Throughout, we adopt 
\begin{equation}
R^{\rho}{}_{\sigma\mu\nu}=\partial_{\mu}\Gamma^{\rho}{}_{\nu\sigma}-\partial_{\nu}\Gamma^{\rho}{}_{\mu\sigma}+\Gamma^{\rho}{}_{\mu\lambda}\Gamma^{\lambda}{}_{\nu\sigma}-\Gamma^{\rho}{}_{\nu\lambda}\Gamma^{\lambda}{}_{\mu\sigma}\,,\label{eq:riemann-convention-app}
\end{equation}
so that 
\begin{equation}
(\nabla_{\mu}\nabla_{\nu}-\nabla_{\nu}\nabla_{\mu})V^{\rho}=R^{\rho}{}_{\sigma\mu\nu}V^{\sigma}.
\end{equation}
With this convention, the geodesic-deviation equation takes the form given below.

\subsection{Covariant derivation}
Consider a smooth two-parameter family of timelike geodesics 
\begin{equation}
    x^{\mu}=x^{\mu}(\tau,s)\,,
\end{equation}
where $\tau$ is proper time along each geodesic and $s$ labels neighboring members of the congruence. We define 
\begin{equation}
    u^{\mu}\coloneqq\frac{\partial x^{\mu}}{\partial\tau}\,,
    \qquad
    \xi^{\mu}\coloneqq\frac{\partial x^{\mu}}{\partial s}\,.
\end{equation}
The vector $u^{\mu}$ is the unit tangent, 
\begin{equation}
    u^{\mu}u_{\mu}=-1,\qquad u^{\nu}\nabla_{\nu}u^{\mu}=0\,,
\end{equation}
whereas $\xi^{\mu}$ is the infinitesimal separation vector.
Since $\tau$ and $s$ are independent coordinates on the two-dimensional surface swept out by the congruence, 
\begin{equation}
    [u,\xi]^{\mu}=0\,,
\end{equation}
since $\partial^2 x^\mu/\partial\tau\partial s=\partial^2 x^\mu/\partial s\partial\tau$. For a torsion-free connection this implies 
\begin{equation}
    \nabla_{u}\xi=\nabla_{\xi}u \,.
\end{equation}
Using the curvature identity together with the geodesic equation, one finds 
\begin{equation}
    \boxed{\frac{D^{2}\xi^{\mu}}{D\tau^{2}}=-R^{\mu}{}_{\alpha\nu\beta}u^{\alpha}u^{\beta}\xi^{\nu}\,.}\label{eq:geo-dev-cov-app}
\end{equation}
Notice that 
\begin{equation}
    \frac{D^{2}\xi^{\mu}}{D\tau^{2}}=u^{\alpha}\nabla_{\alpha}\left(u^{\beta}\nabla_{\beta}\xi^{\mu}\right)=u^{\alpha}u^{\beta}\nabla_{\alpha}\nabla_{\beta}\xi^{\mu}
\end{equation}
only after imposing $u^{\alpha}\nabla_{\alpha}u^{\beta}=0$.

\subsection{Local freely falling frame}
We introduce an orthonormal tetrad $e_{\hat{a}}{}^{\mu}$ along the
reference geodesic, 
\begin{equation}
    g_{\mu\nu}e_{\hat{a}}{}^{\mu}e_{\hat{b}}{}^{\nu}=\eta_{\hat{a}\hat{b}}\,,
    \qquad\eta_{\hat{a}\hat{b}}=\mathrm{diag}(-1,1,1,1) \,,
\end{equation}
with 
\begin{equation}
    e_{\hat{0}}{}^{\mu}=u^{\mu},\qquad\nabla_{u}e_{\hat{a}}{}^{\mu}=0\,.
\end{equation}
The frame is therefore freely falling and non-rotating. Choosing simultaneous separations in the local rest frame, $\xi^{\hat{0}}=0$, \cref{eq:geo-dev-cov-app} becomes 
\begin{equation}
    \boxed{\frac{\text{d}^{2}\xi^{\hat{i}}}{\text{d}\tau^{2}}=-R_{\hat{0}\hat{i}\hat{0}\hat{j}}\xi^{\hat{j}}}\,.\label{eq:geo-dev-tetrad-app}
\end{equation}
The symmetric tensor 
\begin{equation}
    \mathcal{E}_{\hat{i}\hat{j}}\coloneqq R_{\hat{0}\hat{i}\hat{0}\hat{j}}
\end{equation}
is the electric part of the Riemann tensor, namely the physical tidal
matrix measured by the observer.

\subsection{Gauge-invariant scalar variables}\label{app:gauge-inv-variables}
For the scalar sector, we assume that the scalar spatial gauge has already been fixed by setting $E=0$. The perturbed metric \cref{eq:FLRW-unitary} can then be written as 
\begin{align}
        ds^{2}={} & -(1+2\psi)\text{d}t^{2}+2\,\partial_{i}\chi\,\text{d}t\,\text{d}x^{i}+a^{2}(1+2\zeta)\delta_{ij}\text{d}x^{i}\text{d}x^{j}\,,
\end{align}
where the background lapse has been set to unity. With this choice, the gauge-invariant combinations mentioned in the main text are 
\begin{align}
    \mathcal{A} & =\psi-\frac{\text{d}}{\text{d}t}\left(\frac{\zeta}{H}\right) \,,\\
    \mathcal{B} & =\chi+\frac{\zeta}{H} \,.
\end{align}
For later comparison with Ref.~\cite{Palomares:2026iwt}, it is useful to record the dictionary between gauge-invariant variables. After imposing our scalar spatial gauge $E=0$, their metric variables obey $A_{\rm PZK}=\psi$, $\psi_{\rm PZK}=\zeta$, and $aB_{\rm PZK}=\chi$. Their gauge-invariant combinations then map to ours as
\begin{align}
    \Psi_{\rm PZK}&=-\Phi, &
    \Phi_{\rm PZK}&=\Psi, \\
    \mathcal A_{\psi,\rm PZK}&=\mathcal A, &
    \mathcal B_{\rm PZK}&=\mathcal B, \label{eq:PZK-GI-dictionary-app}
\end{align}
where $\mathcal B_{\rm PZK}=a(B_{\rm PZK}+a\dot E_{\rm PZK})+\psi_{\rm PZK}/H$ is the variable denoted $\mathcal B$ in Ref.~\cite{Palomares:2026iwt}. Their gauge-invariant variable $\chi_{\rm GI}^{\rm PZK}$ instead satisfies
\begin{equation}
    \chi_{\rm GI}^{\rm PZK}
    =\mathcal B+\frac{\mathcal R}{H}
    =\frac{\mathcal R-\Phi}{H},
    \label{eq:PZK-chiGI-app}
\end{equation}
where $\mathcal R=-\zeta+H\delta\phi/\dot{\bar\phi}$ in our sign convention, so $\mathcal R_{\rm PZK}=-\mathcal R$. Hence $H\chi_{\rm GI}^{\rm PZK}$ is dimensionless, and in unitary gauge ($\delta\phi=0$) one has $\chi_{\rm GI}^{\rm PZK}=\chi$.
In the following, we use the gauge-invariant Bardeen potentials $\Phi$ and $\Psi$ (defined in \cref{eq:Phi_Bardeen_unitary,eq:Psi_Bardeen_unitary}, respectively) to describe the relative motion of neighboring geodesics. Having already imposed $E=0$, the Newtonian gauge is obtained by further setting $\chi=0$, so that the metric reads 
\begin{equation}
    ds^{2}=-(1+2\Psi)\text{d}t^{2}+a^{2}(1-2\Phi)\delta_{ij}\text{d}x^{i}\text{d}x^{j} \,.\label{eq:newtonian-metric-app}
\end{equation}

For the scalar field that defines the background clock, a kinematic gauge-invariant curvature perturbation is the curvature on uniform-field slices,
\begin{align} \label{eq:calR-definition}
    {\cal R} \coloneqq -\zeta+H\frac{\delta\phi}{\dot\phi} \,.
\end{align}
This definition follows directly from the scalar gauge transformations and does not use the gravitational field equations. Hence, in unitary gauge, $\delta\phi=0$, one has the exact identity ${\cal R}=-\zeta$. In a minimally coupled Einstein--scalar system the constraint equations allow one to rewrite the same quantity in the familiar Bardeen-potential form
\begin{equation}
    {\cal R}=\Phi-\frac{\mathcal H(\Phi'+\mathcal H\Psi)}{\mathcal H'-\mathcal H^2}\,.
\end{equation}
In this one formula primes denote conformal-time derivatives, with $\mathcal H=aH$. The latter expression, and analogous GR relations involving the uniform-density curvature perturbation, should not be imported unchanged into a general Horndeski background because the scalar constraint equations are modified. None of our stability conclusions relies on that GR-specific rewriting.
The scalar part of the shear (in the unitary gauge) is given by
\begin{align}\label{eq:shear-tensor}
    \sigma_{ij} = -\left(\partial_i\partial_j -\frac13 \nabla^2 \delta_{ij}\right) \chi \,,
\end{align}
where $\sigma_{ij}$ has coordinate spatial indices and refers to the normal congruence of the uniform-field slices. Its orthonormal components are $\sigma_{\hat i\hat j}=a^{-2}\sigma_{ij}$ and therefore scale as $(k^2/a^2)\chi$. This is the local shear amplitude; the potential $\chi$ by itself is not an orthonormal tensor component.

\subsection{Tidal tensor in Newtonian gauge}
For a nearly comoving reference geodesic, the peculiar velocity is already first order. The boost from the Newtonian-gauge orthonormal frame to the freely falling frame changes $R_{\hat{0}\hat{i}\hat{0}\hat{j}}$ only beyond linear order, because the mixed background components vanish. Hence, at first order, one may evaluate the tidal tensor using
\begin{equation}
    e_{\hat{0}}=(1-\Psi)\partial_{t} \,,
    \qquad e_{\hat{i}}=\frac{1+\Phi}{a}\partial_{i} \,.
\end{equation}
One obtains 
\begin{align}
    R_{\hat{0}\hat{i}\hat{0}\hat{j}}={} & -\frac{\ddot{a}}{a}\delta_{ij}+\left[\ddot{\Phi}+2H\dot{\Phi}+H\dot{\Psi}+2\frac{\ddot{a}}{a}\Psi\right]\delta_{ij}\nonumber \\
    & +\frac{1}{a^{2}}\partial_{i}\partial_{j}\Psi \,.\label{eq:tidal-newtonian-app}
\end{align}
The first term is the isotropic FLRW tidal field. The square bracket contains a spatially isotropic correction, while the Hessian term has both trace and trace-free parts. This expression is evaluated at Newtonian coordinate time $t_N$. Although its perturbation is expressible in Bardeen variables, comparing measurements to the background at a fixed scalar-field clock or fixed observer proper time requires an additional background-time displacement whenever $H^2+\dot H$ varies.

\subsection{\textit{E}-fold time}
In this appendix we define 
\begin{equation}
    \mathcal{N}\coloneqq\ln\left(\frac{a}{a_{{\rm ini}}}\right)\,,
    \qquad\mathcal{H}_{\tau}\coloneqq\frac{\text{d}\mathcal{N}}{\text{d}\tau}\,,
    \qquad\epsilon\coloneqq-\frac{\dot{H}}{H^{2}}=-\frac{H'}{H} \,.
\end{equation}
Here, a prime denotes a total derivative with respect to $\mathcal{N}$ along the reference geodesic, 
\begin{equation}
    X'\coloneqq\frac{\text{d}X}{\text{d}\mathcal{N}} \,.
\end{equation}
It should be emphasized that this is a reparameterization of the reference geodesic, namely $\mathcal{N}=\mathcal{N}\bigl(t(\tau)\bigr)$, where $\text{d}\tau^{2}=-ds^{2}$, whereas $\epsilon$ is a purely background quantity. Then 
\begin{equation}
    \frac{\text{d}^{2}X}{\text{d}\tau^{2}}=\mathcal{H}^{2}_{\tau}\left[X''+\frac{\text{d}\ln\mathcal{H}_{\tau}}{\text{d}\mathcal{N}}X'\right] \,.
\end{equation}
For a nearly comoving geodesic in Newtonian gauge, assuming the peculiar proper velocity $v^{i}=a\,\text{d}x^{i}/\text{d}t=\mathcal{O}(\delta)$, where $\delta$ represents the perturbative order,\footnote{We are neglecting the velocity contribution $(1-2\Phi)\delta_{ij}v^{i}v^{j}$, which is of second perturbative order, in the exact expression for $\text{d}t/\text{d}\tau$.} we have 
\begin{equation}
    \frac{\text{d}t}{\text{d}\tau}=1-\Psi\,,
    \qquad\mathcal{H}_{\tau}=H(1-\Psi) \,,
\end{equation}
so that 
\begin{equation}
    \frac{\text{d}\ln\mathcal{H}_{\tau}}{\text{d}\mathcal{N}}=\frac{H'}{H}-\Psi'\,,
    \qquad\frac{1}{\mathcal{H}^{2}_{\tau}}=\frac{1+2\Psi}{H^{2}} \,.
\end{equation}
The geodesic-deviation equation becomes 
\begin{align}
    \boxed{\xi^{\hat{i}\,\prime\prime}-(\epsilon+\Psi')\,\xi^{\hat{i}\,\prime}=\left[1-\epsilon-\Phi''-(2-\epsilon)\Phi'-\Psi'\right]\xi^{\hat{i}}-\frac{1}{a^{2}H^{2}}\partial^{i}\partial_{j}\Psi\,\xi^{\hat{j}}} \,.\label{eq:geo-dev-N-app}
\end{align}
The explicit term proportional to $\Psi$ cancels between the tidal tensor and the conversion from proper time to $\mathcal N$. This is a reparameterization identity, not a proof that an arbitrary homogeneous lapse perturbation produces no tidal change. A physical comparison still requires the same clock and initial conditions in the perturbed and background geometries.

\subsection{Physical versus comoving separation}\label{app:physical-vs-comoving}
In exact FLRW, 
\begin{equation}
    R_{\hat{0}\hat{i}\hat{0}\hat{j}}=-\frac{\ddot{a}}{a}\delta_{ij} \,,
\end{equation}
and therefore 
\begin{equation}
    \frac{\text{d}^{2}\xi^{\hat{i}}}{\text{d}t^{2}}=\frac{\ddot{a}}{a}\xi^{\hat{i}} \,.
\end{equation}
The solution $\xi^{\hat{i}}\propto a$ is simply the Hubble expansion of a physical distance and is not an instability.
As an intermediate coordinate-time quantity, define the separation normalized by the background scale factor at $t_N$, 
\begin{equation}
    q^{\hat{i}}\coloneqq\frac{\xi^{\hat{i}}}{a} \,.
\end{equation}
\Cref{eq:geo-dev-N-app} then becomes 
\begin{align}
    \boxed{q^{\hat{i}\,\prime\prime}+(2-\epsilon-\Psi')q^{\hat{i}\,\prime}+\left[\Phi''+(2-\epsilon)\Phi'\right]q^{\hat{i}}+\frac{1}{a^{2}H^{2}}\partial^{i}\partial_{j}\Psi\,q^{\hat{j}}=0} \,.\label{eq:q-dev-N-app}
\end{align}
In the unperturbed limit, 
\begin{equation}
    q^{\hat{i}\,\prime\prime}+(2-\epsilon)q^{\hat{i}\,\prime}=0 \,,
\end{equation}
which admits $q^{\hat{i}}=\mathrm{const.}$

\subsection{Clock-matched tidal response}
Let $A_b=H^2+\dot H$, so the background tidal matrix is $\bar{\mathcal E}_{\hat i\hat j}=-A_b\delta_{ij}$. Using \cref{eq:Phi_Bardeen_unitary,eq:Psi_Bardeen_unitary}, the gradient-free Newtonian tidal correction is
\begin{align}\label{eq:tidal-isotropic-unitary}
    \delta\mathcal E^N_{\rm iso}
    &=H\dot\psi+2A_b\psi-\ddot\zeta-2H\dot\zeta-\dot A_b\chi.
\end{align}
The $\dot\chi$ and $\ddot\chi$ terms cancel identically. The remaining $-\dot A_b\chi$ term is precisely the background-clock displacement between Newtonian and uniform-field slices. Since $\delta\phi_N/\dot\phi=\chi$, the clock-matched tidal perturbation is
\begin{align}
    \delta\mathcal E^{(\phi)}_{\hat i\hat j}
      &\coloneqq\delta\mathcal E^N_{\hat i\hat j}
        -\dot{\bar{\mathcal E}}_{\hat i\hat j}
          \frac{\delta\phi_N}{\dot\phi} \nonumber\\
      &=\left[H\dot\psi+2A_b\psi-\ddot\zeta-2H\dot\zeta\right]\delta_{ij}
        +\frac{1}{a^2}\partial_i\partial_j(\psi+\dot\chi).
      \label{eq:tidal-uniform-clock}
\end{align}
At linear order a change between nearly comoving observer frames does not alter this subtraction, because the relevant mixed FLRW curvature components vanish. The result contains no direct, gradient-free $\chi$ term for \emph{any} background $H(t)$, not only for de Sitter.

For $H=p/t$, the term $-\dot A_b\chi=2p(p-1)\chi/t^3$ in \cref{eq:tidal-isotropic-unitary} has envelope $t^{p/2-3}$ on the oscillatory Weyl branch. It is therefore constant in envelope at $p=6$, but this is a statement about an unmatched Newtonian-time perturbation. It does not establish a physical tidal-instability threshold. The same clock issue applies to linear perturbations of $R$ and $\mathcal K$, whose power-law background values are time-dependent.

An observer-proper-time comparison makes the cancellation equally explicit. Write the proper time along a nearly comoving reference geodesic as
\begin{equation}
    \tau=t_N+\Delta\tau_N(t_N),\qquad
    \dot{\Delta\tau}_N=\Psi.
\end{equation}
After fixing a synchronization condition on an initial physical hypersurface, the tidal perturbation relative to $\bar{\mathcal E}(\tau)$ is
\begin{align}
    \delta\mathcal E^{(\tau)}_{\hat i\hat j}
       &=\delta\mathcal E^N_{\hat i\hat j}
          -\dot{\bar{\mathcal E}}_{\hat i\hat j}\Delta\tau_N,
       \label{eq:tidal-proper-clock}\\
    \Delta\tau_N&=\chi+\int^t\psi(t_1)\,\mathrm{d}t_1+C(\mathbf x),
\end{align}
where the constant $C$ is fixed by that synchronization. Thus
\begin{equation}
    \delta\mathcal E^{(\tau)}_{\hat i\hat j}
       =\delta\mathcal E^{(\phi)}_{\hat i\hat j}
          +\dot A_b\left[\int^t\psi(t_1)\,\mathrm{d}t_1+C\right]\delta_{ij}.
\end{equation}
The direct growing $\chi$ contribution cancels here as well. The remaining lapse integral and the gradient and curvature terms are not set to zero by this identity and must be assessed with the full constrained solution.

To factor out the background Hubble expansion at the same proper time, the appropriate separation is
\begin{equation}
    Q^{\hat i}(\tau)=\frac{\xi^{\hat i}(\tau)}{a(\tau)}.
\end{equation}
To first order, geodesic deviation gives
\begin{equation}
    \frac{\mathrm d^2Q^{\hat i}}{\mathrm d\tau^2}
      +2H(\tau)\frac{\mathrm dQ^{\hat i}}{\mathrm d\tau}
       =-\delta\mathcal E^{(\tau)\hat i}{}_{\hat j}Q^{\hat j}.
    \label{eq:proper-q-deviation}
\end{equation}
By contrast, the quantity $q=\xi/a(t_N)$ used in \cref{eq:q-dev-N-app} satisfies $Q=(1-H\Delta\tau_N)q$ at this order. Its growth can contain a clock displacement. For example, at zeroth spatial-gradient order a pure slicing perturbation with $\Phi=-H\chi$ and $\Psi=\dot\chi$ can give $q\propto1+H\chi$, while $Q$ remains constant. Thus growth of $q$ alone is not a clock-independent tidal effect.

In de Sitter, the remaining direct shear-gradient term decays as $a^{s-2}$ for a branch $\chi\propto a^s$, with $0<\operatorname{Re}s\leq1$ for $\calM^2>0$. For an accelerating power law and $\calM t\gg1$, its oscillatory envelope scales as $(k^2/a^2)\dot\chi\propto t^{-3p/2}$. These decays are consistent with the curvature diagnostics in the main text. They do not establish that all normalized responses remain small or that the integral of the tidal forcing in \cref{eq:proper-q-deviation} stays harmless for arbitrary initial data. In particular, absolute boundedness, smallness relative to a decaying background scale, and bounded integrated geodesic response are distinct criteria. Finite-gradient constraints, nonlinear interactions, the duration of the background phase and its matching to later cosmology remain necessary parts of a complete physical-stability analysis.

\subsection{Electric Weyl response}\label{app:observables}
The electric Weyl tensor measured by a nearly comoving observer has the linear scalar expression
\begin{equation}
 E^{\rm W}_{\hat i\hat j}\equiv C_{\hat0\hat i\hat0\hat j}
 =\frac{1}{2a^2}\left(\partial_i\partial_j-
       \frac{\delta_{ij}}3\nabla^2\right)(\Phi+\Psi).
 \label{eq:observable-electric-Weyl}
\end{equation}
The background Weyl tensor vanishes, so no background-clock subtraction is required at this order; this kinematic relation is also discussed in Ref.~\cite{Li:2015hga}. It is not the full Riemann tide. For a massive de Sitter branch with $(k^2/a^2)\chi\propto a^{\lambda_\pm}$ with $\lambda_\pm=-\frac32\pm\frac12\sqrt{1-\frac{4\calM^2}{H_I^2}}$, its envelope scales as $a^{\operatorname{Re}\lambda_\pm}$ and decays for $\calM^2>0$. For the accelerating power-law branch at $\calM t\gg1$, it scales as $t^{-3p/2}$. These are fixed-$k$ leading linear scalings, not a bound on nonlinear corrections or on an integral over an unspecified initial spectrum. Absolute decay, smallness relative to a changing background and a small integrated detector response are different requirements.

\section{Closed equations for $\Phi$ and $\Psi$}\label{app:closed-eq-Bardeen}
Here we sketch the derivation of the closed fourth-order equations of motion for the gauge-invariant Bardeen potentials $\Phi$ and $\Psi$. Their solutions must satisfy the original coupled constraints, so that the two closed equations do not carry eight independent initial data.
To this end we use the definitions \cref{eq:Phi_Bardeen_unitary,eq:Psi_Bardeen_unitary} to make the following field redefinitions
\begin{align}
    \psi&=\Psi-\dot\chi\,,\\
    \zeta &= -H\chi-\Phi\,,
\end{align}
so as to introduce the Bardeen potentials at the level of the Lagrangian.

\subsection{Equation of motion for $\Psi$}
At this point, we can find the equations of motion for the fields $\chi$, $\Phi$, and $\Psi$, that we name by $E_{\chi}=0$, $E_{\Phi}=0$, $E_{\Psi}=0$, respectively. Then we solve these three equations algebraically for $\chi$, $\dot\chi$, and $\ddot{\chi}$, as
\begin{equation}
    \ddot{\chi}=\mathcal{F}_2(\ddot\Phi,\dot\Phi,\Phi,\dot\Psi,\Psi)\,,\quad
    \dot{\chi}=\mathcal{F}_1(\ddot\Phi,\dot\Phi,\Phi,\dot\Psi,\Psi)\,,\quad
    \chi =\mathcal{F}_0(\ddot\Phi,\dot\Phi,\Phi,\dot\Psi,\Psi)\,.
\end{equation}

At this point we construct a constraint equation, namely
\begin{equation}
    \mathcal{C}_1 \equiv \dot{\mathcal{F}}_1-\mathcal{F}_2=0\,.
\end{equation}
This constraint is going to include terms in $\dddot{\Phi}$, so that we solve $\mathcal{C}_1=0$ for $\dddot{\Phi}$, as
\begin{equation}
    \mathcal{C}_1=0\,,\quad{\rm or}\quad \dddot{\Phi}=\tilde{\mathcal{F}}_3(\ddot\Phi,\dot\Phi,\Phi,\ddot\Psi,\dot\Psi,\Psi)\,.
\end{equation}
At this point, we can find a second constraint by imposing
\begin{equation}
    \mathcal{C}_2 \equiv \dot{\mathcal{F}}_0-\mathcal{F}_1=0\,.
\end{equation}
Then we will solve this equation for $\ddot{\Phi}$. In fact, on replacing $\dddot{\Phi}$ with $\tilde{\mathcal{F}}_3$ in $\mathcal{C}_2=0$, we get
\begin{equation}
    \ddot\Phi=\tilde{\mathcal{F}}_2(\dot\Phi,\Phi,\dot\Psi,\Psi)\,.
\end{equation}
We can now build a new constraint as follows,
\begin{equation}
    \mathcal{C}_3 \equiv \dot{\tilde{\mathcal{F}}}_2-\tilde{\mathcal{F}}_3=0\,,
\end{equation}
where we replace the terms in $\dddot{\Phi}$ and $\ddot{\Phi}$ with $\tilde{\mathcal{F}}_3$ and $\tilde{\mathcal{F}}_2$ respectively. This equation can be solved for $\dot\Phi$, to find
\begin{equation}
\dot\Phi=\tilde{\mathcal{F}}_1(\Phi,\ddot\Psi,\dot\Psi,\Psi)\,.
\end{equation}
Once more, we build up another constraint
\begin{equation}
    \mathcal{C}_4 \equiv \dot{\tilde{\mathcal{F}}}_1-\tilde{\mathcal{F}}_2=0\,,
\end{equation}
which can be solved for $\Phi$, once we substitute any of its time derivatives. We find that
\begin{equation}
    \Phi=\tilde{\mathcal{F}}_0(\dddot\Psi,\ddot\Psi,\dot\Psi,\Psi)\,.
\end{equation}
The last step consists of finding the constraint
\begin{equation}
\mathcal{C}_5 \equiv \dot{\tilde{\mathcal{F}}}_0-\tilde{\mathcal{F}}_1=0\,,
\end{equation}
which, after replacing $\Phi$ and its time derivatives, gives the wanted closed equation of motion for the field $\Psi$, as
\begin{equation}\label{eq:closed-eq-Psi}
    \mathcal{C}_5=\mathcal{C}_5(\ddddot\Psi,\dddot\Psi,\ddot\Psi,\dot\Psi,\Psi)=0\,.
\end{equation}

\subsection{Equation of motion for $\Phi$}
In this case, we can use the equation of motion $E_\Psi=0$ to solve it for $\Psi$, as 
\begin{equation}
    \Psi=\Psi(\chi,\dot\chi,\Phi,\dot\Phi)\,.
\end{equation}
and we can replace this equation directly into the Lagrangian density, \cref{eq:action_scalars}. At this point, we can find the equations of motion for $\chi$ and $\Phi$. We name them as $\tilde{E}_{\chi}=0$ and $\tilde{E}_{\Phi}=0$ respectively. We now solve them algebraically for $\dot\chi$ and $\ddot\chi$, finding
\begin{equation}
    \ddot\chi=\mathcal{G}_2(\chi,\Phi,\dot\Phi,\ddot\Phi)\,,\quad
    \dot\chi=\mathcal{G}_1(\chi,\Phi,\dot\Phi,\ddot\Phi)\,.
\end{equation}
We can now find a constraint $\mathcal{D}_1$ as
\begin{equation}
    \mathcal{D}_1\equiv \dot{\mathcal{G}}_1-\mathcal{G}_2=0\,,
\end{equation}
which, after replacing the expressions for the time-derivatives of $\chi$, can be solved for $\chi$ as
\begin{equation}
    \chi=\mathcal{G}_0(\Phi,\dot\Phi,\ddot\Phi,\dddot\Phi)\,.
\end{equation}
We can build up a second constraint, $\mathcal{D}_2$ defined as
\begin{equation}
    \mathcal{D}_2\equiv \dot{\mathcal{G}}_0-\mathcal{G}_1=0\,,
\end{equation}
which corresponds to a closed equation for $\Phi$, after replacing $\chi$ and its time derivatives, of the form
\begin{equation}\label{eq:closed-eq-Phi}
    \mathcal{D}_2=\mathcal{D}_2(\ddddot\Phi,\dddot\Phi,\ddot\Phi,\dot\Phi,\Phi)=0\,.
\end{equation}

We do not display the coefficients or the resulting equations explicitly, as they are lengthy. The analytical solutions of \cref{eq:closed-eq-Psi,eq:closed-eq-Phi} in the superhorizon limit are more involved than those for $\chi$ and $\zeta$ given in the main text, but exhibit the same late-time behavior and stability properties.

\section{Two-scalar reduction at finite momentum}\label{app:two-scalar-normalization}
In this appendix, we derive the kinetic matrix of the reduced scalar system at finite momentum and the leading-order long-wavelength system used in \cref{sec:scalar-perturbations}. Throughout, $k\neq0$; the exactly homogeneous limit is a different constraint problem.

\paragraph{Finite-momentum kinetic matrix.}
After eliminating the lapse in \cref{eq:action_scalars} by means of \cref{eq:psi-solution}, we write the kinetic part of the reduced Lagrangian as $\mathcal L_{s,\rm kin}^{(2)}=(a^3/2)\,\dot{\mathbf x}^{\rm T}\mathbf K_k\dot{\mathbf x}$ with $\mathbf x^{\rm T}=(\zeta,\chi)$. Defining $\mathcal W_k\coloneqq4\alpha k^4/(3a^4)$, we find
\begin{align}
    \mathbf K_k&=
    \begin{pmatrix}
    -6\calGH+\dfrac{18\Theta^2}{\mathcal W_k-\Sigma}
    &-\dfrac{6\mathcal W_k\Theta}{\mathcal W_k-\Sigma}\\[2mm]
    -\dfrac{6\mathcal W_k\Theta}{\mathcal W_k-\Sigma}
    &\dfrac{2\mathcal W_k\Sigma}{\mathcal W_k-\Sigma}
    \end{pmatrix} \,,
    \qquad
    \det\mathbf K_k=-\frac{12\mathcal W_k\left(\calGH\Sigma+3\Theta^2\right)}{\mathcal W_k-\Sigma} \,.
    \label{eq:scalar-K-finite-k}
\end{align}
In the UV limit $|\mathcal W_k|\gg|\Sigma|$, this reduces to the kinetic matrix $\mathbf K$ of \cref{sec:scalar-perturbations}. On the branch $\alpha<0$, $\calGH<0$ and $\calGH\Sigma+3\Theta^2<0$, one has $\Sigma>0$ and $\mathcal W_k<0$, and $\mathbf K_k$ is positive definite for every $k>0$. In the opposite limit, completing the square in $\dot\zeta$ gives the effective kinetic coefficient of the shear,
\begin{equation}
    K_{\chi|\zeta}\equiv K_{22}-\frac{K_{12}^2}{K_{11}}\simeq-\frac{8\alpha}{3}\frac{k^4}{a^4}
    \qquad (k\to0) \,.
    \label{eq:shear-schur}
\end{equation}
The factor $k^2/a^2$ in the normalized shear $\mathfrak s=(k^2/a^2)\chi$ thus follows from the kinetic term itself, rather than from a choice of variable made after inspecting the growing solution.

\paragraph{Leading-order long-wavelength system.}
We define
\begin{equation}
    r_k\coloneqq\frac{k^2}{a^2} \,,\qquad
    \mathfrak s=r_k\chi \,,\qquad
    D_\zeta\coloneqq\calGH+\frac{3\Theta^2}{\Sigma} \,.
\end{equation}
Using $\dot r_k=-2Hr_k$ and eliminating the lapse, the action \cref{eq:action_scalars} at leading order in gradients becomes
\begin{align}
    \psi=-\frac{\Theta}{\Sigma}\left(3\dot\zeta+\mathfrak s\right) \,,\qquad
    \frac{\mathcal L_{s,0}^{(2)}}{a^3}
    =-3D_\zeta\left(\dot\zeta+\frac{\mathfrak s}{3}\right)^2
    +\frac{\calGH}{3}\mathfrak s^2
    -\frac{4\alpha}{3}\left(\dot{\mathfrak s}+H\mathfrak s\right)^2 \,.
    \label{eq:leading-geometric-action}
\end{align}
The momentum conjugate to $\zeta$, $P_\zeta=-6a^3D_\zeta(\dot\zeta+\mathfrak s/3)$, is conserved, and the equations of motion read
\begin{align}
    \dot P_\zeta&=0 \,,\qquad
    \dot\zeta=-\frac{\mathfrak s}{3}-\frac{P_\zeta}{6a^3D_\zeta} \,,\qquad
    \psi=\frac{\Theta P_\zeta}{2a^3\Sigma D_\zeta} \,,\nonumber\\
    \ddot{\mathfrak s}+3H\dot{\mathfrak s}+\left(\calM^2+2H^2+\dot H\right)\mathfrak s
    &=-\frac{P_\zeta}{8\alpha a^3} \,.
    \label{eq:leading-system}
\end{align}
The four independent data are the two amplitudes of the massive shear branches, $P_\zeta$, and the constant mode of $\zeta$. For $P_\zeta=0$ and after separating the constant-curvature solution, the massive branches satisfy $\dot\zeta=-\mathfrak s/3$, and their leading lapse vanishes. In de Sitter, solutions $\propto a^\lambda$ of this system obey the characteristic equation \cref{eq:scalar-characteristic}, whose nonzero roots have negative real parts for $\calM^2>0$ (with $a^{-3/2}\ln a$ at the repeated root $4\calM^2=H_I^2$). Both $\zeta$ and $\mathfrak s$ are therefore bounded for arbitrary data of the leading-order system. For power-law backgrounds, $D_\zeta$ is constant, $\Theta/\Sigma\propto t$, and the lapse sourced by $P_\zeta$ scales as $t^{1-3p}$. Terms such as $r_k\zeta$, which are subleading in this action, have to be restored when they provide the leading shear of the constant-curvature branch.

\paragraph{Bardeen potentials.}
In terms of these variables, the Bardeen potentials \cref{eq:Phi_Bardeen_unitary,eq:Psi_Bardeen_unitary} are given by \cref{eq:Bardeen-reconstruction}. The inverse factor $1/r_k=a^2/k^2$ converts the decaying massive branches of $\mathfrak s$ into the growing branches of $\Phi$ and $\Psi$. A rescaled variable such as $\mathfrak s$, or its canonically normalized counterpart, can therefore remain bounded while the Bardeen potentials, with their fixed metric normalization, grow. These are different representations of the same linear solution, not different stability verdicts.

\bibliography{references}

\end{document}